\documentstyle[12pt,graphicx,subfigure]{article}
\def\Journal#1#2#3#4{{#1} {\bf #2}, #3 (#4)}

\def\NPB{{\em Nucl. Phys.} B}
\def\PLB{{\em Phys. Lett.}  B}
\def\PRL{\em Phys. Rev. Lett.}
\def\PRD{{\em Phys. Rev.} D}

\def\beq{\begin{equation}}
\def\eeq{\end{equation}}

\def\gmin2{(g-2)_\mu}

\catcode`\@=11 

\def\lsim{\mathrel{\mathpalette\@versim<}}
\def\gsim{\mathrel{\mathpalette\@versim>}}
\def\@versim#1#2{\vcenter{\offinterlineskip
    \ialign{$\m@th#1\hfil##\hfil$\crcr#2\crcr\sim\crcr } }}

\def\PRL{Phys. Rev. Lett.}

\def\beq{\begin{equation}}
\def\eeq{\end{equation}}
\def\beqn{\begin{eqnarray}}
\def\eeqn{\end{eqnarray}}
\def\Journal#1#2#3#4{{#1} {\bf #2}, #3 (#4)}

\def\NPB{{\em Nucl. Phys.} B}
\def\PLB{{\em Phys. Lett.}  B}
\def\PRL{\em Phys. Rev. Lett.}
\def\PRD{{\em Phys. Rev.} D}

\def\lsim{\ ^<\llap{$_\sim$}\ }
\def\gsim{\ ^>\llap{$_\sim$}\ }
\def\r2{\sqrt 2}
\def\beq{\begin{equation}}
\def\eeq{\end{equation}}
\def\beqn{\begin{eqnarray}}
\def\eeqn{\end{eqnarray}}

\def\sinW2{\sin^2\theta_W}

\def\mz2{M_{z}^2}
\def\c2b{\cos 2\beta}

\def\mz{M_z}

\def\Fq2{F_{2}(q^2)}
\def\beq{\begin{equation}}
\def\eeq{\end{equation}}

\def\gmin2{(g-2)_\mu}

\def\sec2w{sec^2\theta_W}

\def\lsim{\ ^<\llap{$_\sim$}\ }
\def\gsim{\ ^>\llap{$_\sim$}\ }
\def\r2{\sqrt 2}
\def\beq{\begin{equation}}
\def\eeq{\end{equation}}
\def\beqn{\begin{eqnarray}}
\def\eeqn{\end{eqnarray}}

\def\sinW2{\sin^2\theta_W}

\def\mz2{M_{z}^2}
\def\c2b{\cos 2\beta}

\def\mz{M_z}

\def\Fq2{F_{2}(q^2)}
\def\sq2{\sqrt{2}}

\def\sec2w{sec^2\theta_W}
\begin{document}

\begin{titlepage}

\begin{center}
{\large {~\bf SUSY QCD and SUSY Electroweak Loop Corrections to $b, t$ and
$\tau$ Masses Including the Effects of CP Phases}}\\
\vskip 0.5 true cm
\vspace{2cm}
\renewcommand{\thefootnote}
{\fnsymbol{footnote}}
 Tarek Ibrahim$^{a,b}$ and Pran Nath$^{b}$  
\vskip 0.5 true cm
\end{center}

\noindent
{a. Department of  Physics, Faculty of Science,
University of Alexandria,}\\
{ Alexandria, Egypt\footnote{: Permanent address }}\\ 
{b. Department of Physics, Northeastern University,
Boston, MA 02115-5000, USA} \\
\vskip 1.0 true cm
\centerline{~ Abstract}
\medskip
We compute supersymmetric QCD and supersymmetric electroweak corrections
 to the b and t quark masses and to the $\tau$ lepton
mass in the presence of CP phases valid for all $\tan\beta$. The analysis 
includes  one loop diagrams arising from the exchange of gluinos, charginos 
and neutralinos. We find that the CP effects can change both the magnitude 
as well as the sign of the supersymmetric loop correction.
In the region of parameter space studied we found that
the numerical size of the correction to the b quark mass can be as large as
30 percent and to the $\tau$ lepton mass as much as 5 percent.
 For the top quark mass the correction is typically less than 
 a percent. These corrections are of importance in unified models since 
 $b-\tau$ and $b-t-\tau$ unification is strongly affected  by the size 
 of these corrections. More generally these results will be of importance
 in analyses of quark-lepton textures with the inclusion of CP phases.
  Further, the analysis presented here is
 relevant in a variety of low energy phenomena where supersymmetric
 QCD and supersymmetric electroweak corrections on b mass enter prominently
 in B physics, e.g.,
  in processes such as $B^0_{s,d}\rightarrow l^+\l^-$. 
\end{titlepage}

\section{Introduction}
The b quark mass plays a very important role in theories of 
fundamental interactions as it enters in tests of $b-\tau$ 
unification\cite{arason} and in $b-t-\tau$ unification\cite{shafi}.
[For recent works on Yukawa unification in SU(5) and in SO(10), see 
refs.\cite{baer}].
In addition the b quark mass enters in a variety of low energy 
phenomena. The physical mass, or the pole mass $M_b$, 
 of the b quark $M_b$ is related to the running mass $m_b(M_b)$ by 
 inclusion of QCD corrections and at the two loop level one has\cite{arason}

\beq 
M_b=(1+\frac{4\alpha_3(M_b)}{3\pi}+12.4\frac{\alpha_3(M_b)^2}{\pi^2})
m_b(M_b)
\eeq
Now $m_b(M_b)$ is derived  from $m_b(M_Z)$ by the running of the 
renormalization group equations
and we focus here on the quantity $m_b(M_Z)$  which is the running
$b$ quark mass at the $Z$ scale. This quantity can be written in the
form 
\beq 
m_b(M_Z)=h_b(M_Z)\frac{v}{\sqrt 2}\cos\beta(1+\Delta_b) 
\eeq
Here $h_b(M_Z)$ is the Yukawa coupling for the b quark at the scale 
$M_Z$, $\beta$ is defined so that
$\tan\beta = <H_2>/<H_1>$  where $H_2$ is the Higgs field that gives mass to
  the up quark and $H_1$ is the Higgs field that  gives mass to the down 
  quark and the lepton, and 
 $\Delta_b$ is loop correction to $m_b$. 
 In previous analyses supersymmetric QCD and supersymmetric electroweak
 loop 
corrections to the running b quark mass at the Z boson mass scale have been 
computed and it is found that in the large $\tan\beta$ region these
corrections can be rather 
large\cite{Hall:1993gn,carena94,Pierce:1996zz,carena2000,carena2002}. 
In this paper we investigate the effects of CP phases SUSY QCD 
and SUSY electroweak loop corrections  
on the b quark mass. We then extend the analysis to include the
effects of CP phases on the loop corrections to the
top quark mass and on the loop corrections on the $\tau$ lepton
mass.  It is well known that
supersymmetric theories contain  CP  phases
via the soft SUSY breaking parameters which are in general complex
and thus introduce new sources of CP violation in the theory above
and beyond what is present in the standard model. Thus, for example, 
in mSUGRA\cite{msugra}
the soft SUSY breaking is characterized by the parameters
$m_0$, $m_{\frac{1}{2}}$, $A_0$ and $B_0$, where $m_0$ is the 
universal scalar mass, $m_{\frac{1}{2}}$ is the universal gaugino mass,
$A_0$ is the universal trilinear coupling and $B_0$ is the bilinear
coupling. In addition mSUGRA  contains a parameter $\mu_0$ 
which is the co-efficient of the Higgs mixing term in the 
superpotential. After spontaneous breaking of the electroweak symmetry 
it is convenient to replace $B_0$ by $\tan\beta$ and the thus mSUGRA
at low energy can be described by the parameters $m_0$, $m_{\frac{1}{2}}$, 
$A_0$ and $\tan\beta$.
In the presence of CP violation mSUGRA contains two CP violating phases
which cannot be removed by field redefinitions. These can be chosen to
be the phases of $\mu_0$ and $A_0$, i.e., 
$\mu_0= |\mu_0| e^{i\theta_{\mu}}$
and $A_0 =|A_0| e^{i\alpha_A}$. In the more general scenario of
the minimal supersymmetric standard model (MSSM) one has many more
phases. In this analysis we will consider this more general situation
of MSSM where in general we will allow for independent phases
for the gaugino masses $\tilde m_i$ so that 
\beq
\tilde m_i = | \tilde m_i| e^{i\xi_i}, ~~i=1,2,3
\eeq
where i=3,2,1 refer to the $SU(3)_C$, $SU(2)_L$ and $U(1)$ 
gauge sectors.  
One problem that must be taken into account  with inclusion of CP 
 phases  is that of the current experimental constraints
on the electric dipole moments (edms) of the electron, of the neutron and of the $H_g^{199}$ 
atom. Thus the current limit on the electron edm is 
$d_e< 4.3\times 10^{-27}ecm$\cite{eedm} and  on the neutron  is
$ d_n<6.5\times 10^{-26} ~ecm$\cite{nedm}). Similarly the limit on
the $H_g^{199}$  atom is also very stringent, i.e.,  
$d_{H_g}<9\times 10^{-28} ecm$\cite{atomic}. Typically, with
phases  O(1) the SUSY contributions to the electron and the
neutron edms are already in excess of the current experimental
limits. There are many suggestions on how to overcome this
problem. These include fine tuning\cite{ellis}, 
large sparticle masses\cite{na}, 
 suppression in the context of Left-Right models\cite{bdm2}, 
 the cancellation
 mechanism\cite{incancel} and suppression by the choice of phases only in the
 third generation\cite{chang}.  Specifically in Refs.\cite{incancel}
 it is shown that in a large class of SUSY, string and brane models
 one can have cancellations and the satisfaction of the edm constraints
 for the electron and the neutron. In Ref.\cite{inbrane} it was also
 shown  that if cancellations occur at a point in the parameter space
 then this point can be promoted into a trajectory where the 
 cancellations occur by using a scaling.  
 Further, in Refs.\cite{olive,inhg199}
 the  implementation of the cancellation mechanism including the 
 $H_g^{199}$ atomic constraint has also been carried out. 
 In some of the above scenarios it is then 
 possible to have large phases and also satisfy the edm constraints.
 It is now well known that if the phases are large they affect a variety
 of low energy phenomena. Some recent works in this direction have
 included the effects of CP phases on the neutral Higgs boson 
 system\cite{pilaftsis,inhiggs,Carena:2001fw}, on collider 
 physics\cite{kane,barger,zerwas}, on g-2\cite{ing2}, in 
 $B^0_{s,d}\rightarrow l^+l^-$\cite{inhg199,huang} and in other B 
 physics\cite{masiero1}.
 The number of such phenomena is rather large and a more complete list
 can be found in Ref.\cite{insusy02}. 
 In the analysis below we will assume that the phases are indeed
 large and that the satisfaction of the edm constraints occurs
 because of one of the mechanisms discussed above. 

	The outline of the rest of the paper is as follows: In Sec.2 we 
	discuss the procedure for the computation of the
	SUSY QCD and SUSY electroweak correction to the fermion masses
	and compute explicitly the corrections to the b quark mass 
	arising from the gluino, chargino and neutralino exchange loops.
	The analysis given is valid not just for large $\tan\beta$ but
	for all $\tan\beta$. In Sec.2 we also discuss the limit when 
	$\tan\beta$ is large and the phases vanish and show that our analysis
	in this  limit is in agreement with the previously derived results.
	In Sec.3 we discuss the SUSY QCD and SUSY electroweak corrections
	to the top quark mass with the inclusion of phases arising from
	the gluino, chargino and neutralino exchange loops. In Sec.4
	we give the analysis for the SUSY electroweak correction to the 
	$\tau$ lepton mass arising from the exchange of charginos and 
	neutralinos. As in the case of the analysis of the correction to the
	b quark mass, the corrections to the top mass and the $\tau$ lepton
	mass include all allowed CP phases
	and the analysis is valid not just for all $\tan\beta$. 
	 The analysis given in Secs.2, 3, and
	4 is the first complete one loop analysis of the 
	SUSY QCD and SUSY electroweak corrections to the b,t and $\tau$ 
	lepton masses with inclusion of phases and without any apriori 
	limitation on $\tan\beta$. The analytic analysis of Secs.2, 3 and 
	4 contain the main new results of this paper. In Sec.5 we give a 
	numerical analysis of the size of the loop corrections to the
	b, t and $\tau$ lepton masses. Sec.6 is devoted to 
	summary and conclusions.

\section{Correction to the b Quark Mass}
We begin by exhibiting the general technique  we use in 
the computation of the corrections to the b quark mass.
Our procedure here is similar to that of Ref\cite{carena2000,carena2002}. 
In this procedure
one first evaluates the loop corrections to the Yukawa couplings 
to the Higgs boson. At the tree level the b quark couples to the
neutral component of the $H_1$ Higgs  boson (i.e.,  $H_1^0$)
while the couplings to the $H_2$ Higgs boson is absent. 
Loop corrections produce a shift in the $H_1^0$ coupling as  
expected and in addition also generate a non-vanishing
effective  coupling with $H_2^0$.   Thus the b quark coupling with
the Higgs in the presence of loop corrections can be 
written as\cite{carena2000,carena2002} 
\beqn
-{L}_{bbH^0}= (h_b+\delta h_b) \bar b_R b_L H_1^0 + 
\Delta h_b \bar b_R b_L H_2^{0*} + H.c.
\eeqn
where we are using the normalization on the Higgs fields of 
Ref.\cite{gunion}.
The correction to the b quark mass is then given by 
\beqn
\Delta_b= [\frac{Re (\Delta h_b)}{h_b} \tan\beta 
+\frac{Re \delta h_b}{h_b} ]
\eeqn
\begin{figure}
\hspace*{-0.6in}
\centering
\includegraphics[width=9cm,height=4cm]{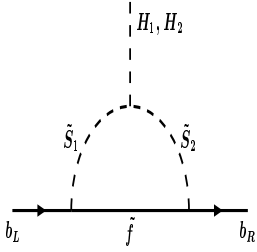}
\caption{ Exhibition of a generic supersymmetric loop contribution to the
 b quark mass. All particles in the loop are heavy supersymmetric partners with
$\tilde S_1,\tilde S_2$ being  heavy scalars and $\tilde f$  a heavy fermion.}
\label{oneloop}
\end{figure}
We now return to the details of the analysis.
In Fig.~\ref{oneloop} we exhibit the typical supersymmetric loop correction, which is either a 
supersymmetric QCD correction or a supersymmetric electroweak correction,
that contributes to the b quark mass. 
 The basic integral that enters in the computation of $\Delta h_b$ and
  $\delta h_b$ is  
  \beq
 I = \int \frac{d^4k}{(2\pi)^4} \frac{m_{\tilde f}+ \not k}
 {(-m_{\tilde f}^2+k^2) (-m_{\tilde S_1}^2+k^2)(-m_{\tilde S_2}^2+k^2)}
 \eeq
 In the approximation that the external momentum can be set to zero the 
 $\not k$ term in the numerator inside the integral in Eq.(6)
  can be neglected and in this case the integral is given by 

\beq
I= \frac{m_{ \tilde f}^2}{(4\pi)^2}  
f(m_{\tilde f}^2,m_{\tilde S_1}^2,m_{\tilde S_2}^2)
\eeq
where the function $f(m^2,m_i^2,m_j^2)$ is given  by

\beqn
f(m^2,m_i^2,m_j^2)
= \frac {1}{(m^2-m_i^2) (m^2-m_j^2)(m_j^2-m_i^2)}
(m_j^2 m^2 ln\frac{m_j^2}{m^2} 
 +m^2 m_i^2ln\frac{m^2}{m_i^2} +m_i^2 m_j^2 ln\frac{m_i^2}{m_j^2})  
 \eeqn
 for the case $i\neq j$ and  
 \beqn
 f(m^2,m_i^2,m_j^2) =\frac {1}{(m_i^2-m^2)^2} (m^2 ln\frac{m_i^2}{m^2} 
 + (m^2-m_i^2))
\eeqn 
for the case i=j.
 In general one will have gluino, chargino and neutralino exchanges 
in the loops. We compute the contribution from each of these below.

\subsection{Gluino Exchange Contribution}
\begin{figure}
\hspace*{-0.6in}
\centering
\includegraphics[width=9cm,height=4cm]{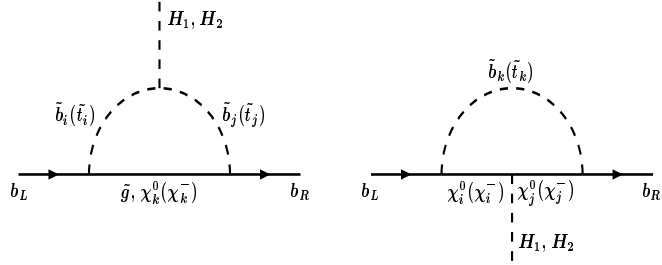}
\caption{One loop contribution to the bottom quark mass involving 
exchange of gluino, charginos and neutralinos in the loop.}
\label{bloop}
\end{figure}

Fig.~\ref{bloop} exhibits the relevant loop involving $\tilde g$ and $\tilde b_{1,2}$
exchanges.
The gluino interaction is given by  

\beqn
-{L}_{\tilde g}= \sqrt 2 g_s  \sum_{\alpha,\beta=1}^{3}
\sum_{a=1}^{8} T^a_{\alpha\beta}
(-\bar b^{\alpha}\frac{1-\gamma_5}{2} \tilde g_a e^{-i\xi_3/2}\tilde b^{\beta}_R 
+ \bar b^{\alpha} \frac{1+\gamma_5}{2}\tilde g_a e^{i\xi_3/2}\tilde b^{\beta}_L )+
 H.c.
\eeqn
where $\alpha, \beta$ are the quark color indices taking on values 1,2,3 and
\beqn
\tilde b_L=\sum_{i=1}^{2} D_{b1i} \tilde b_i,~~~~~ 
\tilde b_R=\sum_{i=1}^{2} D_{b2i} \tilde b_i
\eeqn
Here $D_{bij}$ is the matrix that diagonalizes the b squark mass
matrix  and $\tilde b_i$ are the b squark 
mass eigen states.
For the computation of $\Delta h_b$ we need the $\tilde b\tilde bH_2^0$ 
interaction which is given by 
\beq
{L}_{\tilde{b} \tilde{b}H_2^0}=\sum_{i=1}^2\sum_{j=1}^2 
 G_{ij}\tilde b_i^*\tilde b_j H_2^0 + H.c.
\eeq
 Using the interactions of Eqs.(10) and (12) and the identity 
\beqn
\sum_{a=1}^8 T^a_{\alpha\beta}T^a_{\gamma\delta}=\frac{1}{2} 
[\delta_{\alpha\gamma}\delta_{\beta\delta}
-\frac{1}{3} \delta_{\alpha\beta}\delta_{\gamma\delta} ]
\eeqn
one finds 
\beqn
\Delta h_b = - \sum_{\it i =1}^2 \sum_{j=1}^2 \frac{2\alpha_s}{3\pi} e^{-i\xi_3}m_{\tilde g} 
G_{ij}^* D_{b1i}^{*} D_{b2j} 
f(m_{\tilde g}^2,m_{\tilde b_{\it i}}^2,m_{\tilde b_j}^2)
\eeqn
where 
\beqn
\frac{G_{ij}}{\sqrt 2} =\frac{gM_Z}{2\cos\theta_W} 
\{(-\frac{1}{2} +\frac{1}{3}\sin^2\theta_W)
D^*_{b1i}D_{b1j} -\frac{1}{3} \sin^2\theta_W D^*_{b2i}D_{b2j}\}
\sin\beta\nonumber\\
+\frac{gm_b\mu}{2M_W\cos\beta} D^*_{b1i}D_{b2j}
\eeqn
For the computation of  $\delta h_b$ one needs the $\tilde{b} \tilde{b}
H_1^0$ interaction
which is given by 
\beqn
{L}_{\tilde{b} \tilde{b}H_1^0}=\sum_{i=1}^2\sum_{j=1}^2
 H_{ij} \tilde b_i^*\tilde b_jH_1^0  + H.c.
\eeqn
From Eqs.(10) and (16) one finds 
\beqn
\delta h_b = - \sum_{\it i =1}^2 \sum_{j=1}^2 \frac{2\alpha_s}{3\pi} e^{-i\xi_3}m_{\tilde g} 
H_{ij} D_{b1i}^{*} D_{b2j} 
f(m_{\tilde g}^2,m_{\tilde b_{\it i}}^2,m_{\tilde b_j}^2)
\eeqn
where 
\beqn
\frac{H_{ij}}{\sqrt 2} =-\frac{gM_Z}{2\cos\theta_W} 
\{(-\frac{1}{2} +\frac{1}{3}\sin^2\theta_W)
D^*_{b1i}D_{b1j} -\frac{1}{3} \sin^2\theta_W D^*_{b2i}D_{b2j}\}
\cos\beta\nonumber\\
-\frac{gm_b^2}{2M_W\cos\beta} 
[ D^*_{b1i}D_{b1j} + D^*_{b2i}D_{b2j}]
-\frac{gm_bm_0A_b}{2M_W\cos\beta} 
 D^*_{b2i}D_{b1j} 
\eeqn
\subsection{Chargino Exchange Contribution}
Fig.2 exhibits the relevant loop involving $\chi^-$ and $\tilde t_{1,2}$
exchanges. By carrying a similar analysis of the gluino exchange, the chargino
contribution gives for $\Delta h_b$ the result 
\beqn
\Delta h_b= -\sum_{i=1}^2\sum_{j=1}^2\sum_{k=1}^2    
g^2 E_{ij}^*\{V_{k1}^*D_{t1i}^* -k_t V_{k2}^* D_{t2i}^*\}
(k_b U_{k2}^*D_{t1j})
\frac{m_{\chi_k^+}}{16\pi^2}
f(m_{\chi_k^+}^2,m_{\tilde t_i}^2,m_{\tilde t_j}^2)\nonumber\\
-\sum_{i=1}^2\sum_{j=1}^2\sum_{k=1}^2    
g^2 C_{ij}\{V_{i1}^*D_{t1k}^* -k_t V_{i2}^* D_{t2k}^*\}
(k_b U_{j2}^*D_{t1k})
\frac{m_{\chi_i^+}m_{\chi_j^+}}{16\pi^2}
f(m_{\tilde t_k}^2,m_{\chi_i^+}^2,m_{\chi_j^+}^2)
\eeqn
where 
\beqn
\frac{E_{ij}}{\sqrt 2} =\frac{gM_Z}{2\cos\theta_W} \{(\frac{1}{2} -\frac{2}{3}\sin^2\theta_W)
D^*_{t1i}D_{t1j} +\frac{2}{3} \sin^2\theta_W D^*_{t2i}D_{t2j}\}
\sin\beta\nonumber\\
-\frac{gm_t^2}{2M_W\sin\beta} 
[ D^*_{t1i}D_{t1j} + D^*_{t2i}D_{t2j}]
-\frac{gm_tm_0A_t}{2M_W\sin\beta} 
 D^*_{t2i}D_{t1j} 
\eeqn
and 
\beqn
\frac{C_{ij}}{\sqrt 2}= -\frac{g}{2\sin\beta} 
[ \frac{m_{\chi_i^+}}{2M_W} \delta_{ij}
-Q^*_{ij} \cos\beta - R^*_{ij} ] 
\eeqn
and where
\beqn
Q_{ij}= \sqrt{\frac{1}{2}} U_{i2}V_{j1}\nonumber\\
R_{ij}=\frac{1}{2M_W} [\tilde m_2^* U_{i1} V_{j1} + \mu^* U_{i2} V_{j2}] 
\eeqn
and
\beqn
k_{t(b)}=\frac{m_{t(b)}}{\sqrt 2 m_W \sin\beta (\cos\beta)}
\eeqn
The matrices U and V diagonalize the chargino mass matrix by a
biunitary transformation. Similarly
the matrix $D_{tij}$ diagonalizes the stop mass matrix.\\

\noindent
The computation of  $\delta h_b$ gives  
\beqn
\delta h_b= -\sum_{i=1}^2\sum_{j=1}^2\sum_{k=1}^2    
g^2 F_{ij}\{V_{k1}^*D_{t1i}^* -k_t V_{k2}^* D_{t2i}^*\}
(k_b U_{k2}^*D_{t1j})
\frac{m_{\chi_k^+}}{16\pi^2}
f(m_{\chi_k^+}^2,m_{\tilde t_i}^2,m_{\tilde t_j}^2)\nonumber\\
-\sum_{i=1}^2\sum_{j=1}^2\sum_{k=1}^2    
g^2 K_{ij}\{V_{i1}^*D_{t1k}^* -k_t V_{i2}^* D_{t2k}^*\}
(k_b U_{j2}^*D_{t1k})
\frac{m_{\chi_i^+}m_{\chi_j^+}}{16\pi^2}
f(m_{\tilde t_k}^2,m_{\chi_i^+}^2,m_{\chi_j^+}^2)
\eeqn
where 

\beqn
\frac{F_{ij}}{\sqrt 2} =-\frac{gM_Z}{2\cos\theta_W} \{(\frac{1}{2} -\frac{2}{3}\sin^2\theta_W)
D^*_{t1i}D_{t1j} +\frac{2}{3} \sin^2\theta_W D^*_{t2i}D_{t2j}\}
\cos\beta\nonumber\\
+\frac{gm_t\mu}{2M_W\sin\beta} 
 D^*_{t1i}D_{t2j}
\eeqn
and 

\beqn
\frac{K_{ij}}{\sqrt 2}= -\frac{g}{2} Q_{ji} 
\eeqn

\subsection{Neutralino Exchange Contribution}
Fig.2 exhibits the relevant loop involving neutalinos ($\chi^0_k, k=1-4$) 
 and sbottom ($\tilde b_{1,2}$)  exchanges. The contribution of 
 these loops to $\Delta h_b$ is 
\beqn
\Delta h_b= 
-\sum_{i=1}^2\sum_{j=1}^2 \sum_{k=1}^4 
2G_{ij}^* 
\{\alpha_{bk}D_{b1j}-\gamma_{bk}D_{b2j}\} 
   \{\beta_{bk}^*D_{b1i}^*+\alpha_{bk}D_{b2i}^*\} 
  \frac{m_{\chi_k^0}}{16\pi^2}
f(m_{\chi_k^0}^2,m_{\tilde b_i}^2,m_{\tilde b_j}^2)\nonumber\\
-\sum_{i=1}^4\sum_{j=1}^4 \sum_{k=1}^2 
2\Gamma_{ij} 
\{\alpha_{bj}D_{b1k}-\gamma_{bj}D_{b2k}\} 
   \{\beta_{bi}^*D_{b1k}^*+\alpha_{bi}D_{b2k}^*\} 
  \frac{m_{\chi_i^0} m_{\chi_j^0}}{16\pi^2}
f(m_{\tilde b_k}^2, m_{\chi_i^0}^2, m_{\chi_j^0}^2 )
\eeqn
where
\beqn
\alpha_{bk}=\frac{gm_bX_{3k}}{2m_W \cos\beta}\nonumber\\
\beta_{bk}= eQ_b X^{'*}_{1k} +\frac{g}{\cos\theta_W}X^{'*}_{2k}
(T_{3b}-Q_b \sin^2\theta_W)\nonumber\\
\gamma_{b k} =e Q_b X'_{1k} -\frac{gQ_b\sin^2\theta_W}{\cos\theta_W}
X_{2k}'
\eeqn
and where 
\beqn
X'_{1k}=X_{1k}\cos\theta_W +X_{2k}\sin\theta_W\nonumber\\
X'_{2k}=-X_{1k}\sin\theta_W +X_{2k}\cos\theta_W
\eeqn
Here $T_{3b}=-1/2$, $Q_b=-1/3$ and $e=g\sin\theta_W$ and the matrix X
diagonalizes the neutralino mass matrix (see, e.g., Ref.\cite{inmssm} for
notation and definitions). 
 $\Gamma_{ij}$ appearing in Eq.(27) is defined by 
\beqn
\frac{\Gamma_{ij}}{\sqrt 2}= 
-\frac{g}{2\sin\beta} [ \frac{m_{\chi_i^0}}{2M_W} \delta_{ij}
-Q^{''*}_{ij} \cos\beta
- R^{''*}_{ij}] 
\eeqn
where 
\beqn
gQ^{''}_{ij}= \frac{1}{2} [ X_{3i}^* (gX_{2j}^* -g' X_{1j}^*) +
(i\leftarrow \rightarrow j) ]\nonumber\\
R_{ij}^{''}= \frac{1}{2M_W} [ \tilde m_1^* X_{1i}^*X_{1j}^* 
+ \tilde m_2^* X_{2i}^*X_{2j}^*
-\mu^* (X_{3i}^*X_{4j}^* + X_{4i}^*X_{3j}^*) ] 
\eeqn
Finally the analysis of the same loops  gives the
 neutralino exchange contribution to 
$\delta h_b$ 
\beqn
\delta h_b= 
-\sum_{i=1}^2\sum_{j=1}^2 \sum_{k=1}^4 
2H_{ij} 
\{\alpha_{bk}D_{b1j}-\gamma_{bk}D_{b2j}\} 
   \{\beta_{bk}^*D_{b1i}^*+\alpha_{bk}D_{b2i}^*\} 
  \frac{m_{\chi_k^0}}{16\pi^2}
f(m_{\chi_k^0}^2,m_{\tilde b_i}^2,m_{\tilde b_j}^2)\nonumber\\
-\sum_{i=1}^4\sum_{j=1}^4 \sum_{k=1}^2 
2\Delta_{ij} 
\{\alpha_{bj}D_{b1k}-\gamma_{bj}D_{b2k}\} 
   \{\beta_{bi}^*D_{b1k}^*+\alpha_{bi}D_{b2k}^*\} 
  \frac{m_{\chi_i^0} m_{\chi_j^0}}{16\pi^2}
f(m_{\tilde b_k}^2, m_{\chi_i^0}^2, m_{\chi_j^0}^2 )
\eeqn
where
\beqn
\frac{\Delta_{ij}}{\sqrt 2}= -\frac{g}{2} Q_{ij}{''} 
\eeqn
A similar analysis for the corrections to the top quark mass 
is given in Sec.3 and for the $\tau$ lepton mass is given in
Sec.4.

\subsection{The Limit of Vanishing Phases and Large $\tan\beta$ }
We check the results of Sec.2.1-2 with previous analyses 
by taking the limit when phases vanish and when $\tan\beta$ is large.
 We first look at the gluino exchange contribution. 
Here in the large $\tan\beta$ limit in Eq.(5) the first term 
dominates. Further, setting  $\xi_3 =0$ and using the
approximation of ignoring the sbottom mixing  we find
 that the correction to the b quark mass from the 
gluino exchange  from Eq.(14) is given by 
\beq
\Delta_b^{\tilde g}= \frac{2\alpha_s\mu M_{\tilde g}}
{3\pi}   \tan\beta f(m_{\tilde b_1}^2, m_{\tilde b_2}^2,M_{\tilde g}^2)
\eeq
This is precisely the result obtained in previous analyses for the
case with large $\tan\beta$ and no phase(see Eq. 6 of 
\cite{Hall:1993gn}). 
Next we consider the chargino exchange contribution given by Eq.(19).
To compare the results with
previous analyses we assume  that $|\mu|>>M_W, \tilde m_2$.
In this limit the matrices U and V are unity,
and only the terms i=2, j=1, k=2 in the sum contributes. Further, using
the approximation that the matrix D is unity, putting $m_{\chi_2^+}=\mu$ 
and setting the phases to zero one finds
\beqn
\Delta m_b^{\chi^+} =
\frac{g^2}{h_b}\tan\beta E_{21}
 k_b k_t 
\frac{\mu}{16\pi^2}
f(\mu^2,m_{\tilde t_i}^2,m_{\tilde t_j}^2)
\eeqn
From the definitions of $k_b, k_t$, 
and in the limit of large $\tan \beta$ and having the relation that
$m_0 A_t>> M_Z$ one finds that our chargino result with no phases gives 
\beq
\Delta_b^{\tilde \chi^+}= \frac{g^2 m_t^2  m_0 A_t \mu}
{32 \pi^2 M_W^2 \sin \beta \cos \beta}  f(m_{\tilde t_1}^2, m_{\tilde t_2}^2,\mu^2)
\eeq
This is the result derived in previous
analyses in the large $\tan\beta$ limit in 
the absence of phases(see Eq. 7 of \cite{Hall:1993gn}).
In both of the above cases, i.e., for 
the gluino exchange and for the chargino exchange we find that
the loop correction to the b quark mass increases  with 
$\tan\beta$ and thus it can get large for large $\tan\beta$.

\section{Correction to the top Quark Mass}
In this section we discuss the loop corrections to the top quark
mass in the presence of CP  phases. The analysis proceeds
in a manner very similar to the analysis of the bottom mass correction.
Thus the running of the  top mass at the Z scale is given by 
\beq 
m_t(M_Z)=h_t(M_Z)\frac{v}{\sqrt 2}\sin\beta(1+\Delta_t) 
\eeq
where $\Delta_t$ gives the loop correction to $m_t$. 
The loop correction $\Delta_t$ is then given by  
\beqn
\Delta_t = (\frac{Re (\Delta h_t)}{h_t} cot\beta +
\frac{Re (\delta h_t)}{h_t} )
\eeqn
We carried out a one loop computation of $\Delta h_t$ and find  for
$\Delta h_t$ the result
\beqn
\Delta h_t = - \sum_{\it i =1}^2 \sum_{j=1}^2 \frac{2\alpha_s}{3\pi} e^{-i\xi_3}m_{\tilde g} 
F_{ij}^* D_{t1i}^{*} D_{t2j} 
f(m_{\tilde g}^2,m_{\tilde t_{\it i}}^2,m_{\tilde t_j}^2)\nonumber\\
-\sum_{i=1}^2\sum_{j=1}^2\sum_{k=1}^2    
g^2 H_{ij}^*\{U_{k1}^*D_{b1i}^* -k_b U_{k2}^* D_{b2i}^*\}
(k_t V_{k2}^*D_{b1j})
\frac{m_{\chi_k^+}}{16\pi^2}
f(m_{\chi_k^+}^2,m_{\tilde b_i}^2,m_{\tilde b_j}^2)\nonumber\\
-\sum_{i=1}^2\sum_{j=1}^2\sum_{k=1}^2    
g^2 K_{ji}^*\{U_{i1}^*D_{b1k}^* -k_b U_{i2}^* D_{b2k}^*\}
(k_t V_{j2}^*D_{b1k})
\frac{m_{\chi_i^+}m_{\chi_j^+}}{16\pi^2}
f(m_{\tilde b_k}^2,m_{\chi_i^+}^2,m_{\chi_j^+}^2)\nonumber\\
-\sum_{i=1}^2\sum_{j=1}^2 \sum_{k=1}^4 
2F_{ij}^* 
\{\alpha_{tk}D_{t1j}-\gamma_{tk}D_{t2j}\} 
   \{\beta_{tk}^*D_{t1i}^*+\alpha_{tk}D_{t2i}^*\} 
  \frac{m_{\chi_k^0}}{16\pi^2}
f(m_{\chi_k^0}^2,m_{\tilde t_i}^2,m_{\tilde t_j}^2)\nonumber\\
-\sum_{i=1}^4\sum_{j=1}^4 \sum_{k=1}^2 
2\Delta_{ij}^* 
\{\alpha_{tj}D_{t1k}-\gamma_{tj}D_{t2k}\} 
   \{\beta_{ti}^*D_{t1k}^*+\alpha_{ti}D_{t2k}^*\} 
  \frac{m_{\chi_i^0} m_{\chi_j^0}}{16\pi^2}
f(m_{\tilde t_k}^2, m_{\chi_i^0}^2, m_{\chi_j^0}^2 ) 
\eeqn
where 
\beqn
\alpha_{tk} =\frac{g_2m_tX_{4k}}{2m_W\sin\beta}\nonumber\\
\beta_{tk}=eQ_tX_{1k}^{'*} +\frac{g}{\cos\theta_W} X_{2k}^{'*}
(T_{3t}-Q_t\sin^2\theta_W)\nonumber\\
\gamma_{tk}=eQ_t X_{1k}'-\frac{gQ_t\sin^2\theta_W}{\cos\theta_W}
X_{2k}'
\eeqn
and where $Q_t=\frac{2}{3}$ and $T_{3t}=\frac{1}{2}$. 
Similarly for $\delta h_t$ we find the result
\beqn
\delta h_t = - \sum_{\it i =1}^2 \sum_{j=1}^2 \frac{2\alpha_s}{3\pi} e^{-i\xi_3}m_{\tilde g} 
E_{ij} D_{t1i}^{*} D_{t2j} 
f(m_{\tilde g}^2,m_{\tilde t_{\it i}}^2,m_{\tilde t_j}^2)\nonumber\\
-\sum_{i=1}^2\sum_{j=1}^2\sum_{k=1}^2    
g^2 G_{ij}\{U_{k1}^*D_{b1i}^* -k_b U_{k2}^* D_{b2i}^*\}
(k_t V_{k2}^*D_{b1j})
\frac{m_{\chi_k^+}}{16\pi^2}
f(m_{\chi_k^+}^2,m_{\tilde b_i}^2,m_{\tilde b_j}^2)\nonumber\\
-\sum_{i=1}^2\sum_{j=1}^2\sum_{k=1}^2    
g^2 C_{ji}^*\{U_{i1}^*D_{b1k}^* -k_b U_{i2}^* D_{b2k}^*\}
(k_t V_{j2}^*D_{b1k})
\frac{m_{\chi_i^+}m_{\chi_j^+}}{16\pi^2}
f(m_{\tilde b_k}^2,m_{\chi_i^+}^2,m_{\chi_j^+}^2)\nonumber\\
-\sum_{i=1}^2\sum_{j=1}^2 \sum_{k=1}^4 
2E_{ij} 
\{\alpha_{tk}D_{t1j}-\gamma_{tk}D_{t2j}\} 
   \{\beta_{tk}^*D_{t1i}^*+\alpha_{tk}D_{t2i}^*\} 
  \frac{m_{\chi_k^0}}{16\pi^2}
f(m_{\chi_k^0}^2,m_{\tilde t_i}^2,m_{\tilde t_j}^2)\nonumber\\
-\sum_{i=1}^4\sum_{j=1}^4 \sum_{k=1}^2 
2\Gamma_{ij}^* 
\{\alpha_{tj}D_{t1k}-\gamma_{tj}D_{t2k}\} 
   \{\beta_{ti}^*D_{t1k}^*+\alpha_{ti}D_{t2k}^*\} 
  \frac{m_{\chi_i^0} m_{\chi_j^0}}{16\pi^2}
f(m_{\tilde t_k}^2, m_{\chi_i^0}^2, m_{\chi_j^0}^2 ) 
\eeqn

\section{Correction to the $\tau$ Lepton Mass}
In this section we discuss the loop corrections  to the $\tau$ lepton
mass. The analysis of the loop corrections to the $\tau$ lepton mass
are very similar to the analysis of the b quark mass with one 
important distinction. Unlike the correction to the b quark mass 
where there are contributions arising from the gluino, chargino and
neutralino exchanges, for the case of correction to the $\tau$ lepton
one has contributions arising only from the chargino and neutralino 
exchanges.
 Keeping this distinction in mind we follow closely 
the procedure of the analysis of b quark mass. Thus we begin by 
defining the $\tau$ mass  at the Z scale by  
\beq 
m_{\tau}(M_Z)=h_{\tau}(M_Z)\frac{v}{\sqrt 2}\cos\beta(1+\Delta_{\tau}) 
\eeq
where $\Delta_{\tau}$ is loop correction to $m_{\tau}$.
$\Delta_{\tau}$ is given by 
\beqn
\Delta_{\tau} =[\frac{Re\Delta h_{\tau}}{h_{\tau}} \tan\beta 
+\frac{Re\delta h_{\tau}}{h_{\tau}}]
\eeqn
The remaining loop analysis is very similar to that of the b quark mass
and we omit the details and give the results below. The analysis of
$\Delta h_{\tau}$ gives  
\beqn
\Delta h_{\tau} = 
-\sum_{i=1}^2\sum_{j=1}^2    
g^2 C_{ij}V_{i1}^*  k_{\tau} U_{j2}^*
\frac{m_{\chi_i^+}m_{\chi_j^+}}{16\pi^2}
f(m_{\tilde \nu_{\tau}}^2,m_{\chi_i^+}^2,m_{\chi_j^+}^2)\nonumber\\
-\sum_{k=1}^2
g^2 E_{\tau}^* V_{k1}^*  k_{\tau} U_{k2}^*
\frac{m_{\chi_k^+}}{16\pi^2}
f(m_{\chi_k^+}^2,m_{\tilde \nu_{\tau}}^2,m_{\tilde \nu_{\tau}}^2)\nonumber\\
-\sum_{i=1}^2\sum_{j=1}^2 \sum_{k=1}^4 
2G_{\tau ij}^* 
\{\alpha_{\tau k}D_{\tau 1j}-\gamma_{\tau k}D_{\tau 2j}\} 
   \{\beta_{\tau k}^*D_{\tau 1i}^*+\alpha_{\tau k}D_{\tau 2i}^*\} 
  \frac{m_{\chi_k^0}}{16\pi^2}
f(m_{\chi_k^0}^2,m_{\tilde \tau_i}^2,m_{\tilde \tau_j}^2)\nonumber\\
-\sum_{i=1}^4\sum_{j=1}^4 \sum_{k=1}^2 
2\Gamma_{ij} 
\{\alpha_{\tau j}D_{\tau 1k}-\gamma_{\tau j}D_{\tau 2k}\} 
   \{\beta_{\tau i}^*D_{\tau 1k}^*+\alpha_{\tau i}D_{\tau 2k}^*\} 
  \frac{m_{\chi_i^0} m_{\chi_j^0}}{16\pi^2}
f(m_{\tilde \tau_k}^2, m_{\chi_i^0}^2, m_{\chi_j^0}^2 ) 
\eeqn
where 
\beqn
k_{\tau}=\frac{m_{\tau}}{\sqrt 2 m_W \cos\beta}\nonumber\\ 
\alpha_{\tau k}=\frac{gm_{\tau} X_{3k}}{2m_W \cos\beta} \nonumber\\
\beta_{\tau k} =e Q_{\tau} X_{1k}^{'*} 
+\frac{g}{\cos\theta_W} X_{2k}^{'*}(T_{3\tau} -Q_{\tau} 
\sin^2\theta_W)\nonumber\\
\gamma_{\tau k} =e Q_{\tau} X_{1k}' -\frac{gQ_{\tau}\sin^2\theta_W}
{\cos\theta_W}X_{2k}'
\eeqn
and where
\beqn
\frac{G_{\tau ij}}{\sqrt 2}=\frac{gM_Z}{2\cos\theta_W} \{(-\frac{1}{2} 
+\sin^2\theta_W)
D^*_{\tau 1i}D_{\tau 1j} - \sin^2\theta_W D^*_{\tau 2i}D_{\tau 2j}\}
\sin\beta\nonumber\\
+\frac{gm_{\tau}\mu}{2M_W\cos\beta} D^*_{\tau 1i}D_{\tau 2j}
\eeqn
and $E_{\tau}$ is
\beq
\frac{E_{\tau}}{\sqrt 2}=\frac{g M_Z}{4 \cos \theta_{W}} \sin\beta
\eeq
Similarly for $\delta h_{\tau}$ one  gets 
\beqn
\delta h_{\tau} = 
-\sum_{i=1}^2\sum_{j=1}^2    
g^2 K_{ij}V_{i1}^*  k_{\tau} U_{j2}^*
\frac{m_{\chi_i^+}m_{\chi_j^+}}{16\pi^2}
f(m_{\tilde \nu_{\tau}}^2,m_{\chi_i^+}^2,m_{\chi_j^+}^2)\nonumber\\
-\sum_{k=1}^2
g^2 F_{\tau} V_{k1}^*  k_{\tau} U_{k2}^*
\frac{m_{\chi_k^+}}{16\pi^2}
f(m_{\chi_k^+}^2,m_{\tilde \nu_{\tau}}^2,m_{\tilde \nu_{\tau}}^2)\nonumber\\
-\sum_{i=1}^2\sum_{j=1}^2 \sum_{k=1}^4 
2H_{\tau ij} 
\{\alpha_{\tau k}D_{\tau 1j}-\gamma_{\tau k}D_{\tau 2j}\} 
   \{\beta_{\tau k}^*D_{\tau 1i}^*+\alpha_{\tau k}D_{\tau 2i}^*\} 
  \frac{m_{\chi_k^0}}{16\pi^2}
f(m_{\chi_k^0}^2,m_{\tilde \tau_i}^2,m_{\tilde \tau_j}^2)\nonumber\\
-\sum_{i=1}^4\sum_{j=1}^4 \sum_{k=1}^2 
2\Delta_{ij} 
\{\alpha_{\tau j}D_{\tau 1k}-\gamma_{\tau j}D_{\tau 2k}\} 
   \{\beta_{\tau i}^*D_{\tau 1k}^*+\alpha_{\tau i}D_{\tau 2k}^*\} 
  \frac{m_{\chi_i^0} m_{\chi_j^0}}{16\pi^2}
f(m_{\tilde \tau_k}^2, m_{\chi_i^0}^2, m_{\chi_j^0}^2 ) 
\eeqn
where 
\beqn
\frac{H_{\tau ij}}{\sqrt 2}=-\frac{gM_Z}{2\cos\theta_W} \{(-\frac{1}{2} +\sin^2\theta_W)
D^*_{\tau 1i}D_{\tau 1j} -\sin^2\theta_W D^*_{\tau 2i}D_{\tau 2j}\}
\cos\beta\nonumber\\
-\frac{gm_{\tau}^2}{2M_W\cos\beta} 
[ D^*_{\tau 1i}D_{\tau 1j} + D^*_{\tau 2i}D_{\tau 2j}]
-\frac{gm_{\tau}m_0A_{\tau}}{2M_W\cos\beta} 
 D^*_{\tau 2i}D_{\tau 1j} 
\eeqn
and 
\beq
\frac{F_{\tau}}{\sqrt 2}=-\frac{g M_Z}{4 \cos \theta_{W}} \cos\beta 
\eeq
\section{Discussion of results}
\begin{figure}
\hspace*{-0.6in}
\centering
\includegraphics[width=12cm,height=16cm]{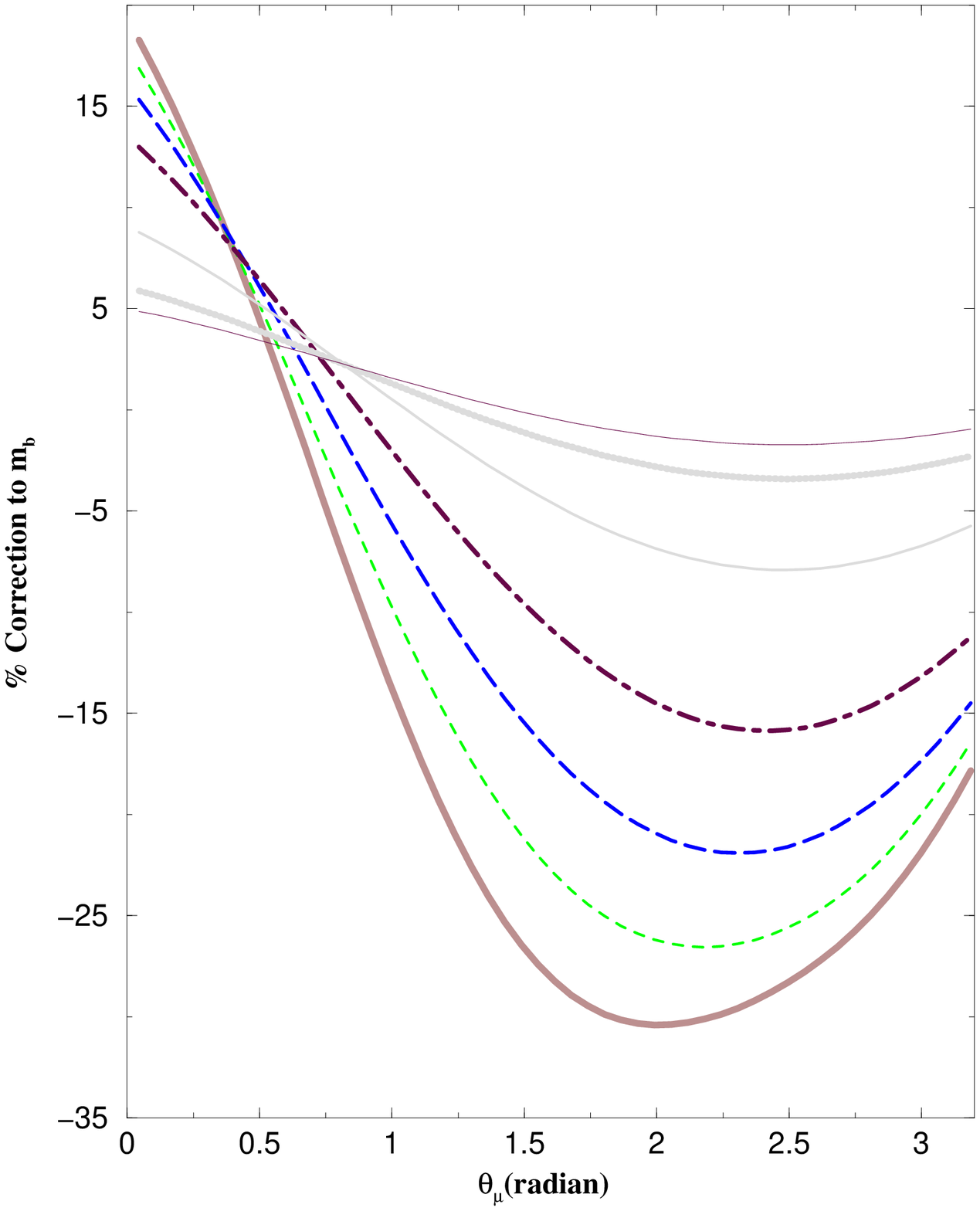}
\caption{ Plot of the b quark mass correction $\Delta m_b/m_b$ 
as a function of $\theta_{\mu}$ 
The curves in descending order at the point
$\theta_{\mu}=\pi$ correspond to 
$\tan\beta =3,5,10,20,30,40,50$. The other input parameters are:  
$m_0=m_{\frac{1}{2}}=200$ GeV,  $\xi_1=.5$, $\xi_2=.659$, 
$\xi_3=.633$, $\alpha_{A_0}=1.0$, and $|A_0|=4$. All angles here 
and in succeeding figure captions are in radians.}
\label{bfigA}
\end{figure}

\begin{figure}
\hspace*{-0.6in}
\centering
\includegraphics[width=12cm,height=16cm]{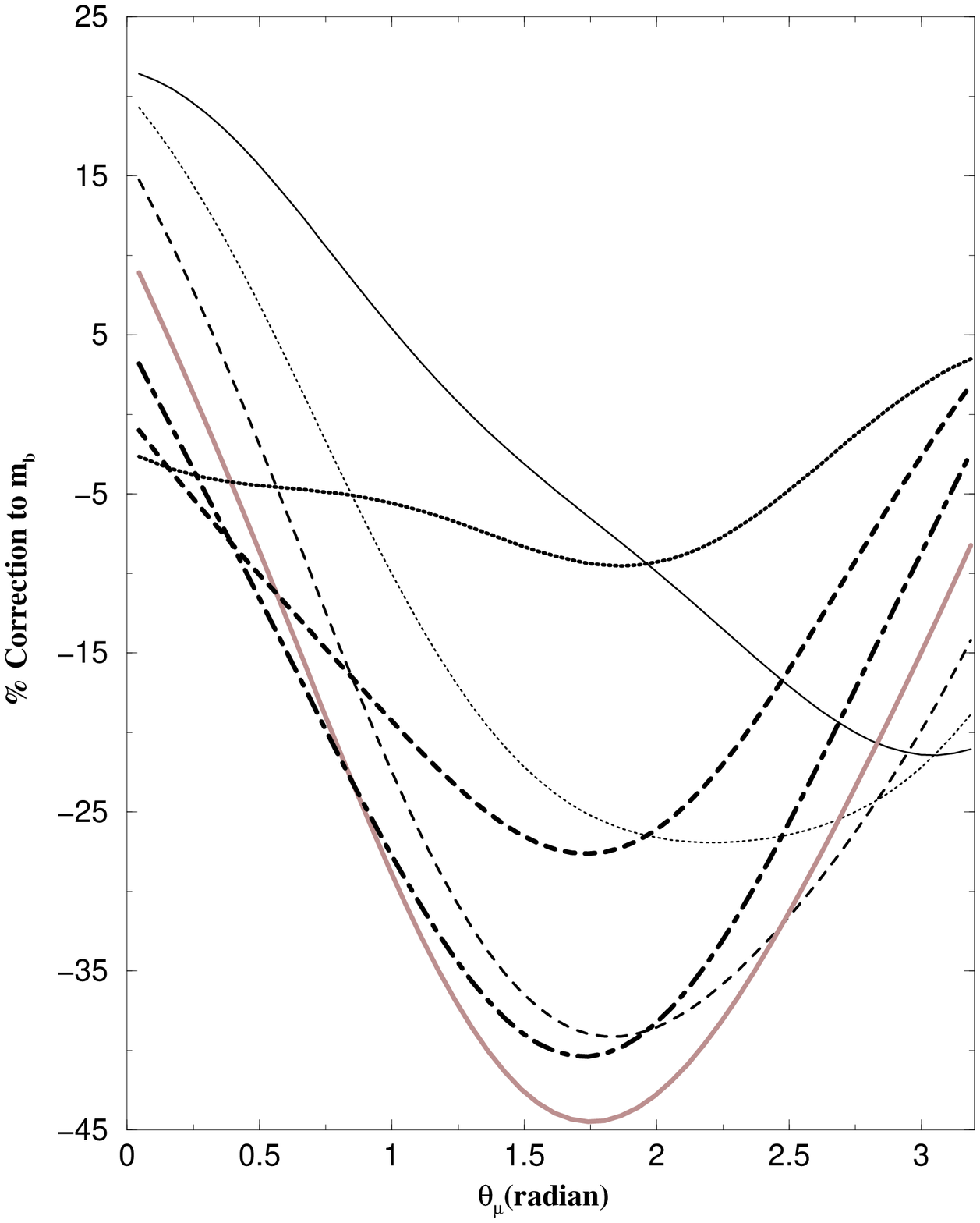}
\caption{ Plot of the b quark mass correction  $\Delta m_b/m_b$ 
 as a function of $\theta_{\mu}$ for various  values of 
$\xi_3$. The other input parameters are:  
$m_0=m_{\frac{1}{2}}=200$ GeV,  $\xi_1=.5$, $\xi_2=.659$,
 $\alpha_{A_0}=1.0$, $|A_0|=4$, and $\tan\beta =50$. The curves 
 in descending order at the point $\theta_{\mu} = 0$ correspond to 
 $\xi_3=0, 0.5,1.0,1.5,2.0,2.5,3.0$. }
\label{bfigB}
\end{figure}

\begin{figure}
\hspace*{-0.6in}
\centering
\includegraphics[width=12cm,height=16cm]{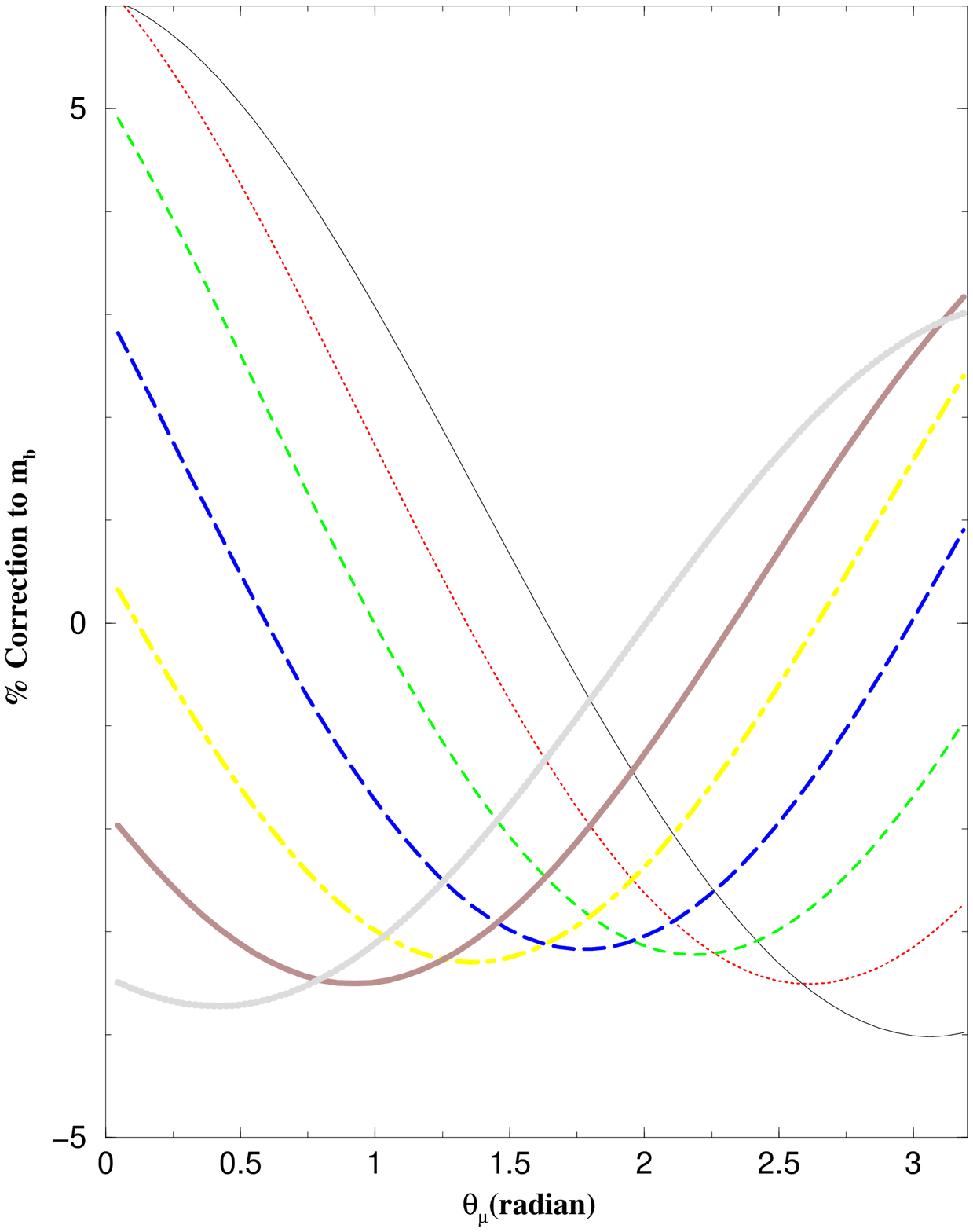}

\caption{  Plot of the b quark mass correction  $\Delta m_b/m_b$ 
 as a function of $\theta_{\mu}$.for various  values of 
$\xi_3$. The other input parameters are:  
$m_0=m_{\frac{1}{2}}=200$ GeV,  $\xi_1=.5$, $\xi_2=.659$,
 $\alpha_{A_0}=1.0$, $|A_0|=4$, and $\tan\beta =5$. The curves 
in descending order at the point $\theta_{\mu} = 0.5$ correspond to  
 $\xi_3 = 0, 0.5,1.0,1.5,2.0,2.5,3.0$. }
\label{bfigC}
\end{figure}

\begin{figure}
\hspace*{-0.6in}
\centering
\includegraphics[width=12cm,height=16cm]{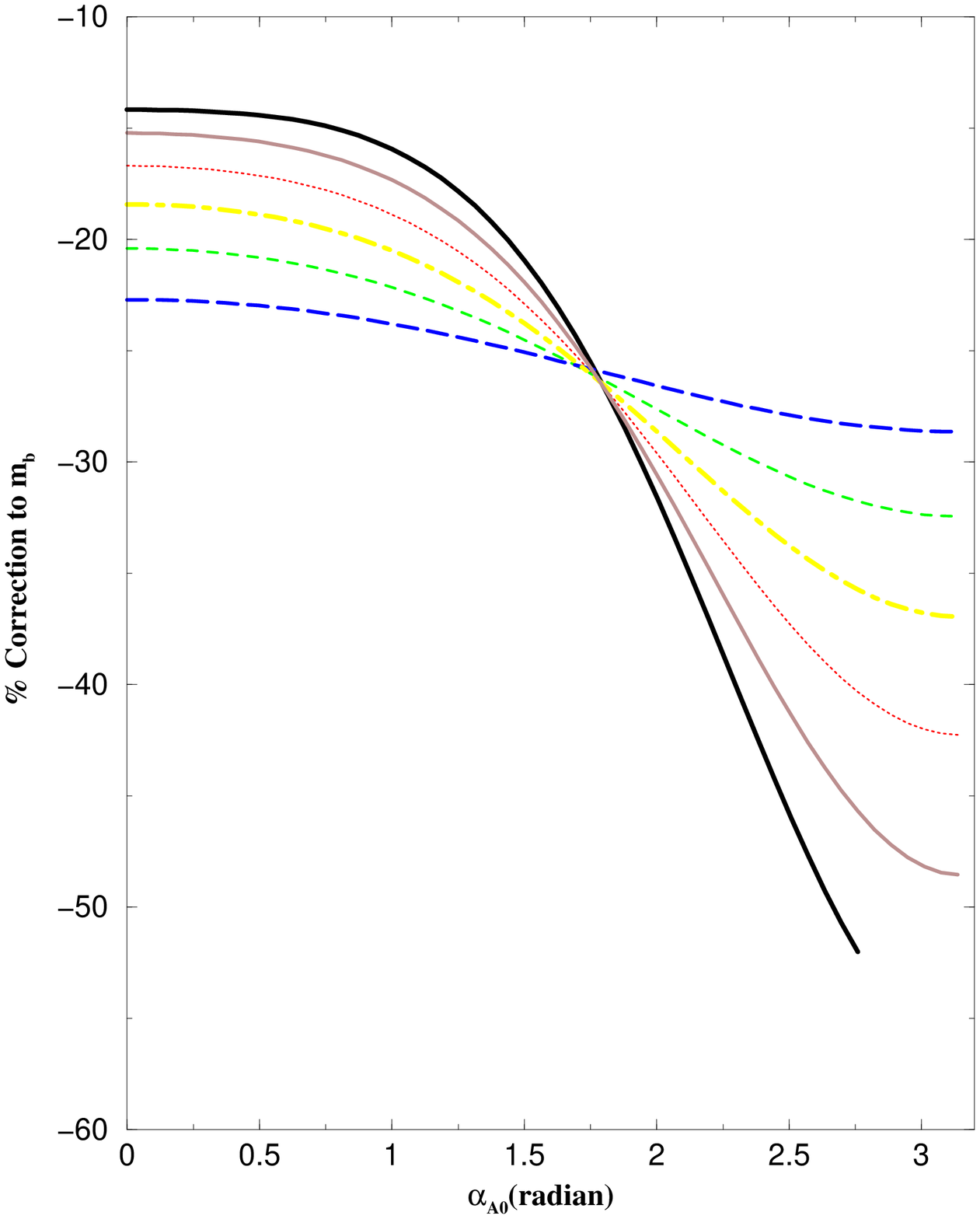}
\caption{Plot of the b quark mass correction  $\Delta m_b/m_b$ 
as a function of $\alpha_{A_0}$ for various values of 
 $|A_0|$.   The other input parameters are:  
$m_0=m_{\frac{1}{2}}=200$ GeV,  $\xi_1=.5$, $\xi_2=.659$, 
$\xi_3=.633$, $\theta_{\mu}=\pi/4$,  and $\tan\beta =50$.
The curves in descending order at the point $ \alpha_{A_0}= 2.5$ 
correspond to $|A_0|=1,2,3,4,5,6$.}
\label{bfigD}
\end{figure}

\begin{figure}
\vspace{-2.5cm}
\hspace*{-0.6in}
\centering
\includegraphics[width=12cm,height=16cm]{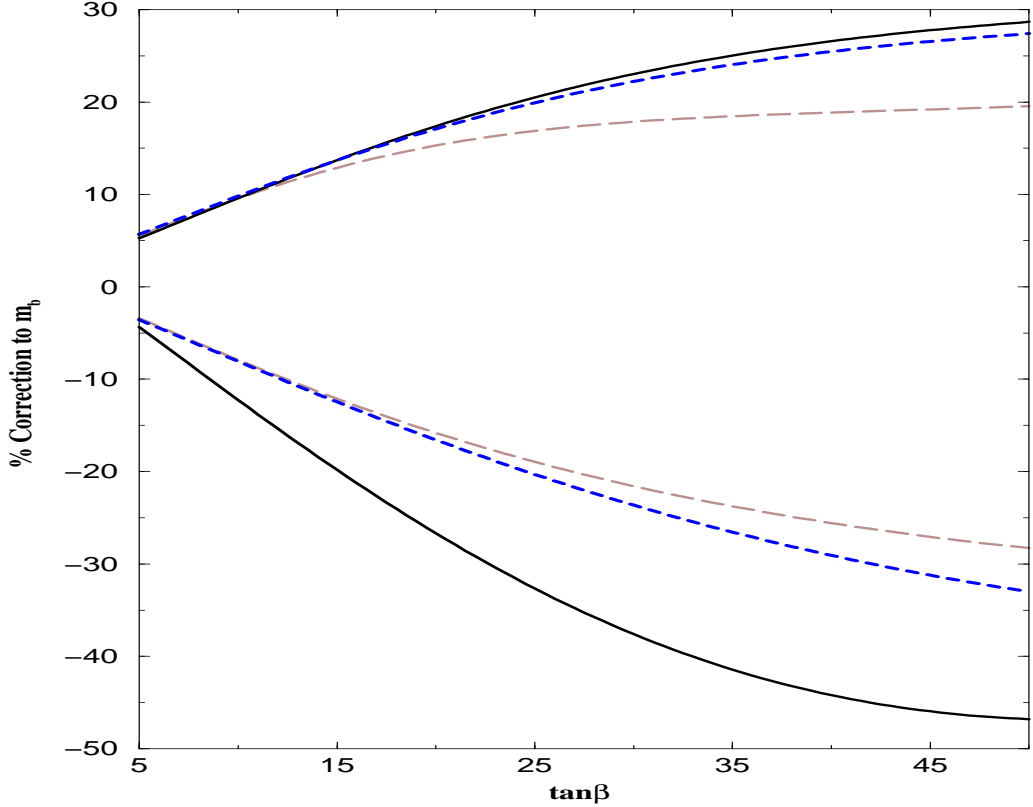}
\caption{Plot of the b quark mass correction  $\Delta m_b/m_b$ 
as a function of $\tan\beta$ for the three
cases (a), (b) and (c) of Table 1 of Ref.\cite{inhg199} where the EDM 
constraints for the electron, the neutron and $H_g^{199}$ atom are satisfied for 
the case $\tan\beta =50$. Among the curves in the lower half plane
corresponding to negative correction to the  b quark mass, the  
long dashed curve corresponds to the case 
(a) where $m_0=m_{\frac{1}{2}}=200$ GeV, $A_0=4$, $\alpha_{A_0}=1$,
  $\xi_1=.5$, $\xi_2=.659$, $\xi_3=.633$; 
  the solid curve corresponds to the case (b) where 
  $m_0=m_{\frac{1}{2}}=370$ GeV, $A_0=4$, $\alpha_{A_0}=2$,
  $\xi_1=.6$, $\xi_2=.653$, $\xi_3=.672$;  
  the dashed curve corresponds to the case (c) where
    $m_0=m_{\frac{1}{2}}=320$ GeV, $A_0=3$, $\alpha_{A_0}=.8$,
  $\xi_1=.4$, $\xi_2=.668$, $\xi_3=.6$. $\theta_{\mu}=2.5$ in all cases.
   The three similar curves in
  the upper half plane are for the three cases above when the phases
  are all set to zero.}
\label{bfigE}
\end{figure}

We begin with a discussion of the effects of CP phases on   
on the SUSY QCD and the SUSY electroweak correction to the b quark mass.
In Fig.~\ref{bfigA} a plot of the b quark mass correction 
 as a function of $\theta_{\mu}$ is given for  values of
 $\tan\beta$ ranging from 3 to 50. One finds that the correction
is very sensitively dependent on $\theta_{\mu}$ as the  value of $\theta_{\mu}$
affects both the sign and the magnitude of the correction. Thus the
correction can vary from zero to as much as  30\% in some regions of
the parameter space and can also change its sign dependent on the 
value of $\theta_{\mu}$. In Fig.~\ref{bfigB} a similar plot is given 
as a function of  $\theta_{\mu}$ for different values of $\xi_3$ for
the case $\tan\beta =50$. One finds that $\Delta m_b$  is also 
a very sensitive function of the phase $\xi_3$ and as in the case of
$\theta_{\mu}$ this phase can 
also change both the magnitude as well as the sign of the b quark
correction.  In Fig.~\ref{bfigC} we give a plot similar to that of
Fig.~\ref{bfigB} for the case of $\tan\beta =5$. The conclusions here
are very similar to those for Fig.~\ref{bfigB} except that the
over all  magnitude of the correction is typically smaller due to
the smaller value of $\tan\beta$ in this case.
In Fig.~\ref{bfigD} a plot of the b quark correction as a function of
$\alpha_{A_0}$ for various values of $|A_0|$ is given. One finds again
a very strong dependence of the b quark correction on both $\alpha_{A_0}$ 
as well as on $|A_0|$. This strong dependence is less obvious than
the dependence on $\theta_{\mu}$ and $\xi_3$ and arises from 
their effects on the stop and sbottom masses and via
the dependence of the matrices that diagonalize the  
stop and sbottom  mass matrices on $\alpha_{A_0}$ and on $|A_0|$.
In Fig.~\ref{bfigE} we give a plot of the b quark correction as
a function of $\tan\beta$ for the three cases of Table 1 of
 Ref.\cite{inhg199} which satisfy the edm constraints for the 
 electron, the neutron and for the $H_g$ atom for value of 
 $\tan\beta =50$. The corresponding cases where the phases are all
 set to zero are also plotted. The plot shows that the effect of 
 phases consistent with the edm constraints can produce large effects
 on the b quark mass.
 
\begin{figure}
\hspace*{-0.6in}
\centering
\includegraphics[width=12cm,height=16cm]{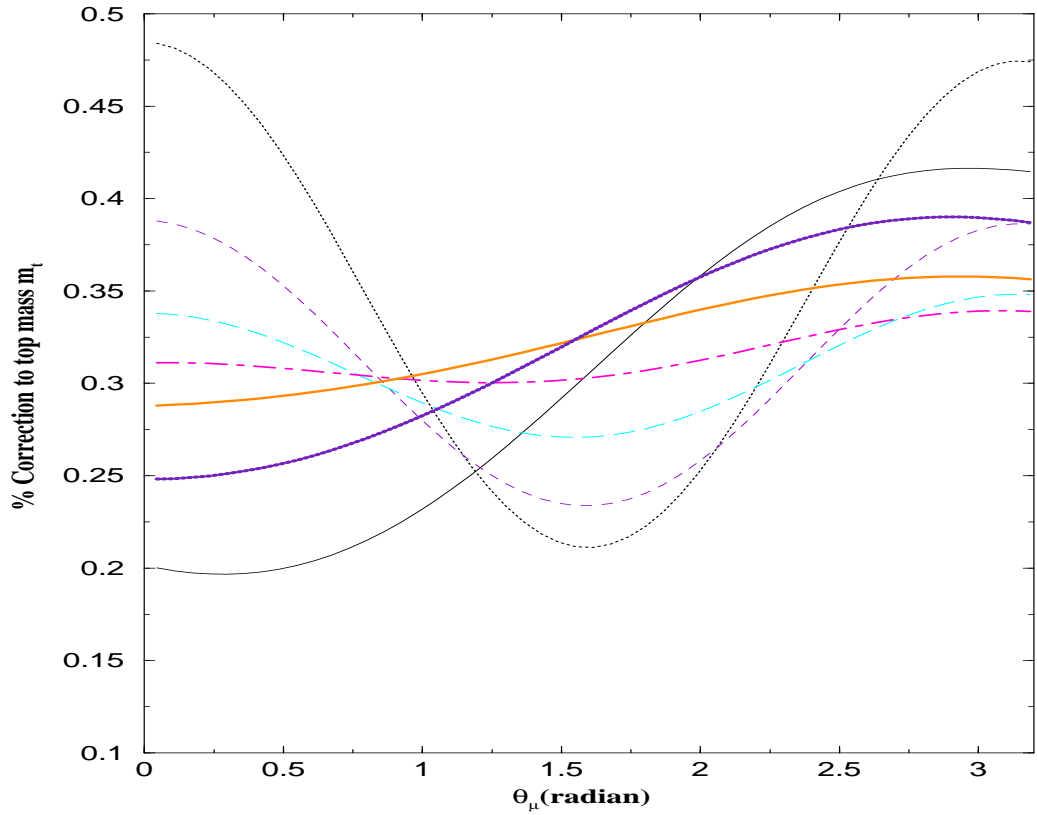}
\caption{Plot of the top quark mass correction $\Delta m_t/m_t$ 
as a function of $\theta_{\mu}$ for values of 
$\tan\beta$ ranging from 3 to 50. The other input parameters are:  
$m_0=m_{\frac{1}{2}}=200$ GeV,  $\xi_1=.5$, $\xi_2=.659$, 
$\xi_3=.633$, $\alpha_{A_0}=1.0$, and $|A_0|=4$.
The curves in ascending order at the point
$\theta_{\mu}= 0$ correspond to $\tan\beta =3,5,10,20,30,40,50$.
 }
\label{tfigA}
\end{figure}

\begin{figure}
\hspace*{-0.6in}
\centering
\includegraphics[width=12cm,height=16cm]{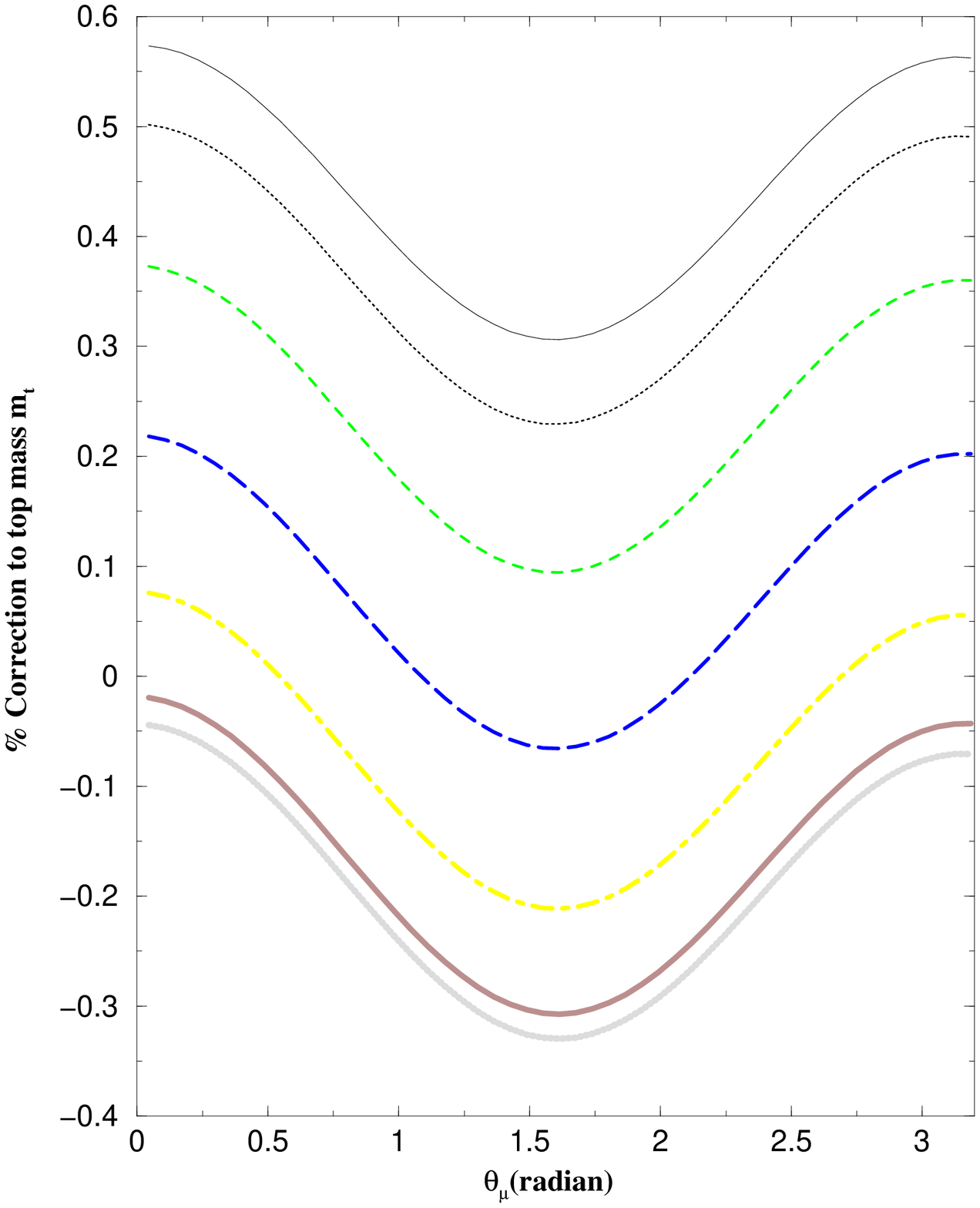}
\caption{Plot of the top quark mass correction $\Delta m_t/m_t$ 
in  percentage as a function of $\theta_{\mu}$ for various  values of 
$\xi_3$. The other input parameters are: 
$m_0=m_{\frac{1}{2}}=200$ GeV,  $\xi_1=.5$, $\xi_2=.659$,
 $\alpha_{A_0}=1.0$, $|A_0|=4$, and $\tan\beta =50$. 
 The curves 
 in descending order at the point $\theta_{\mu} = \pi$ correspond to 
 $\xi_3=0, 0.5,1.0,1.5,2.0,2.5,3.0$.  }
\label{tfigB}
\end{figure}

\begin{figure}
\hspace*{-0.6in}
\centering
\includegraphics[width=12cm,height=16cm]{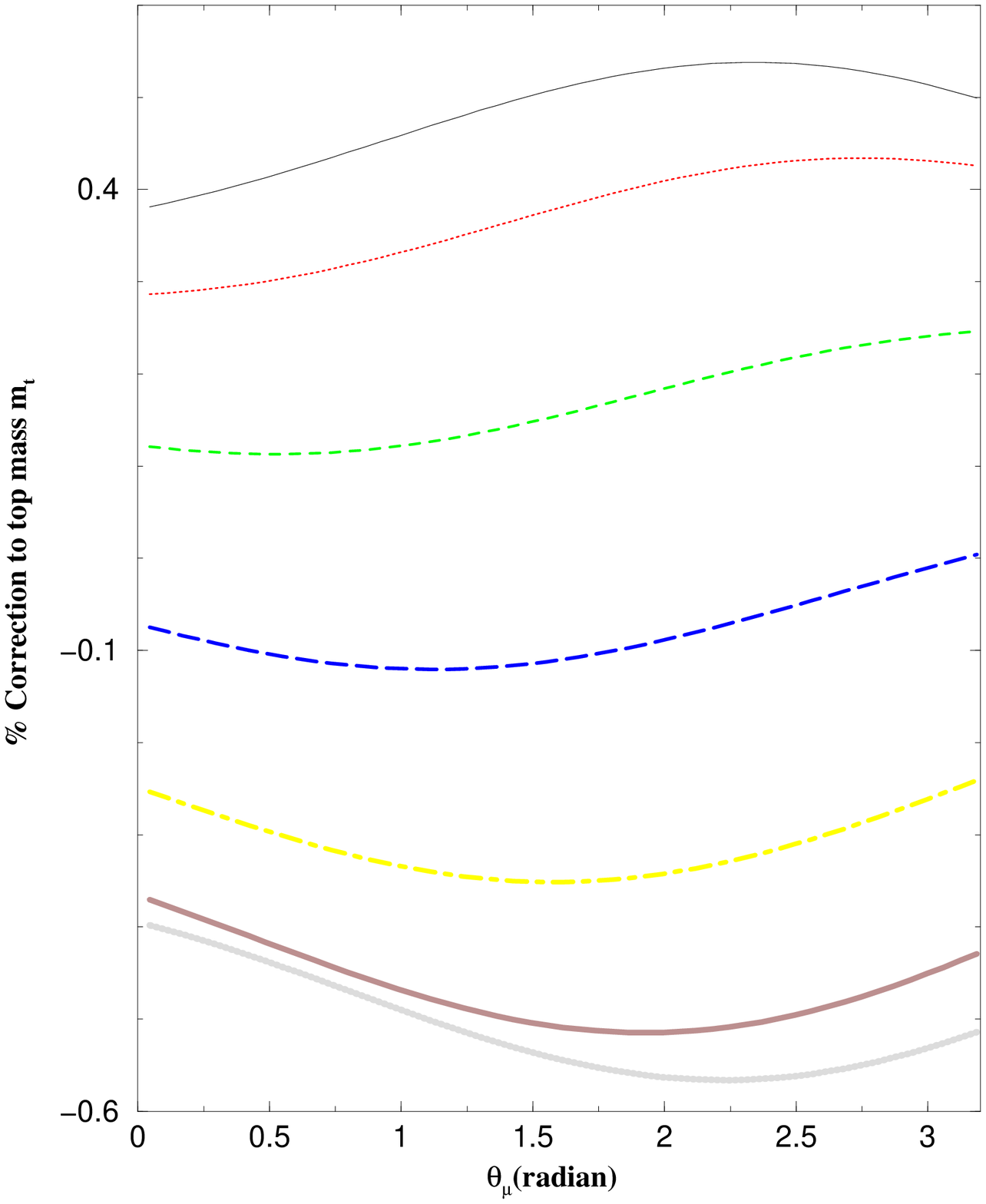}
\caption{Plot of the top quark mass correction $\Delta m_t/m_t$ 
in  percentage as a function of $\theta_{\mu}$ for various  values of 
$\xi_3$. The other input parameters are: 
$m_0=m_{\frac{1}{2}}=200$ GeV,  $\xi_1=.5$, $\xi_2=.659$,
 $\alpha_{A_0}=1.0$, $|A_0|=4$, and $\tan\beta =5$.
  The curves 
 in descending order at the point $\theta_{\mu} = \pi$ correspond to 
   $\xi_3=0, 0.5,1.0,1.5,2.0,2.5,3.0$  }
\label{tfigC}
\end{figure}

\begin{figure}
\hspace*{-0.6in}
\centering
\includegraphics[width=12cm,height=16cm]{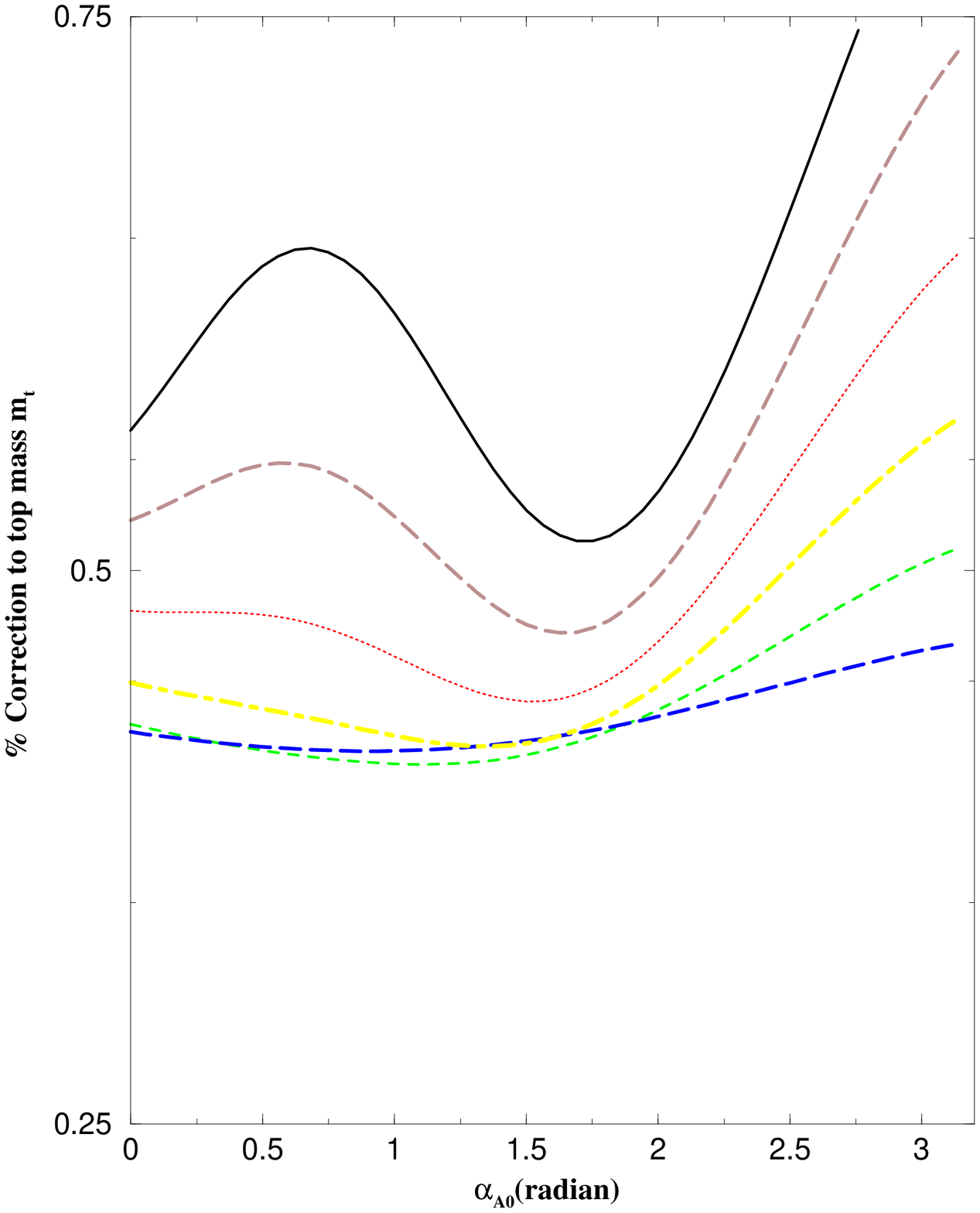}
\caption{Plot of the top quark mass correction $\Delta m_t/m_t$ 
in  percentage as a function of $\alpha_{A_0}$  
for various values of $|A_0|$. The other input parameters are:  
$m_0=m_{\frac{1}{2}}=200$ GeV,  $\xi_1=.5$, $\xi_2=.659$, 
$\xi_3=.633$, $\theta_{\mu}=\pi$ and $\tan\beta =50$.
The curves in ascending order at the point $ \alpha_{A_0}= 2.5$ 
correspond to  $|A_0|=1,2,3,4,5,6$}
\label{tfigD}
\end{figure}

\begin{figure}
\vspace{-2cm}
\hspace*{-0.6in}
\centering
\includegraphics[width=12cm,height=16cm]{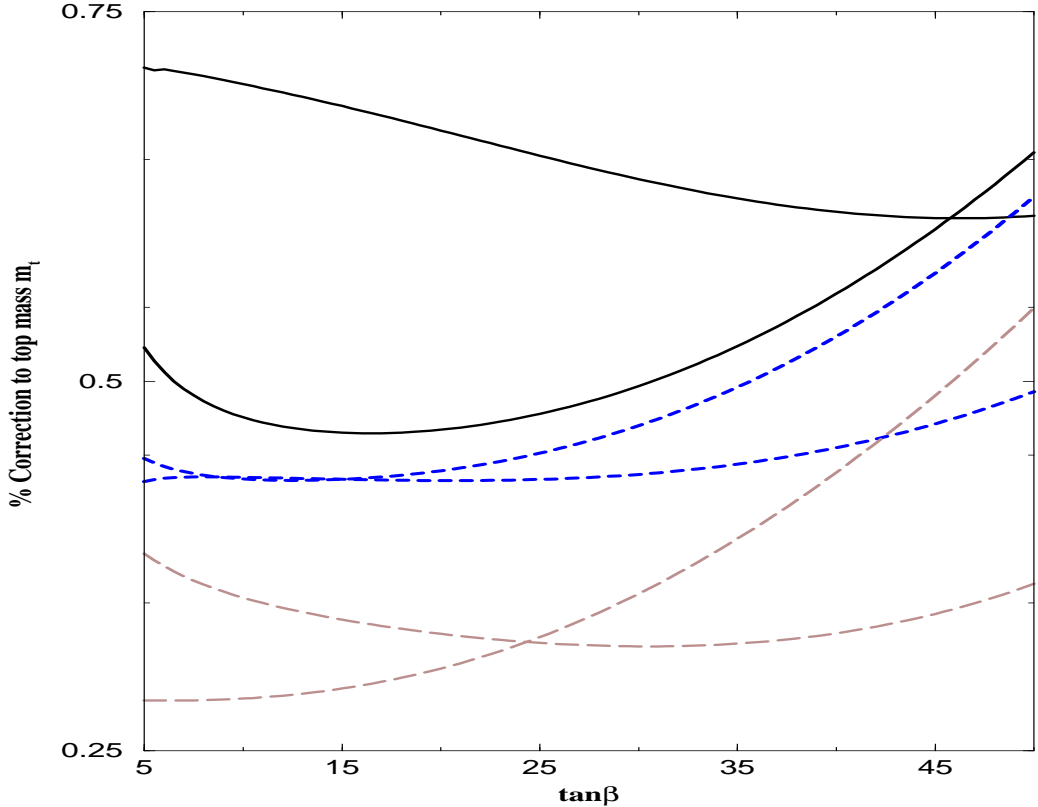}
\caption{Plot of the top quark mass correction $\Delta m_t/m_t$ 
in  percentage as a function of $\tan\beta$ for the three
cases (a), (b) and (c) of Table 1 of Ref.\cite{inhg199} where the EDM 
constraints for the electron, the neutron and $H_g^{199}$ atom are 
satisfied for 
the case $\tan\beta =50$. The long dashed curve with intercept at 
$\tan\beta =50$ of 0.36 is for the case 
(a) where $m_0=m_{\frac{1}{2}}=200$ GeV, $A_0=4$, $\alpha_{A_0}=1$,
  $\xi_1=.5$, $\xi_2=.659$, $\xi_3=.633$.
  The solid  curve with intercept at $\tan\beta =50$ of  0.61
 is for the case (b) where 
  $m_0=m_{\frac{1}{2}}=370$ GeV, $A_0=4$, $\alpha_{A_0}=2$,
  $\xi_1=.6$, $\xi_2=.653$, $\xi_3=.672$. 
  The dashed curve with intercept at $\tan\beta =50$ of 0.49
  is for the case (c) 
    where   $m_0=m_{\frac{1}{2}}=320$ GeV, $A_0=3$, $\alpha_{A_0}=.8$,
  $\xi_1=.4$, $\xi_2=.668$, $\xi_3=.6$. $\theta_{\mu}=2.5$ in all cases.
  The other three similar
  curves have all the same parameters as cases (a), (b), and (c) above
  except that phases are set to zero.}
\label{tfigE}
\end{figure}

  An analysis similar to the above for the
 correction to the top quark mass is given in figures 
 Fig.~\ref{tfigA}, 
 Fig.~\ref{tfigB}, Fig.~\ref{tfigC}, Fig.~\ref{tfigD}, and Fig.~\ref{tfigE}.
  Thus in Fig.~\ref{tfigA} we give a plot of the top quark correction
  as a function of the $\theta_{\mu}$ for all the same parameters
  as in Fig.~\ref{bfigA}. As in the case of the b quark 
  correction here also one finds that the correction to the top 
  quark mass is very strongly dependent on 
  $\theta_{\mu}$. However, unlike the case of the b quark correction
  the correction to the top quark mass is typically less than  
  a percent which means that the  SUSY loop correction  to the top quark 
  mass is likely to be at the level of a GeV or so.
  Such an effect could still be significant at the level of 
  precision theoretical analyses. Very similar conclusions follow
  from the analysis of  
  Fig.~\ref{tfigB}, Fig.~\ref{tfigC}, Fig.~\ref{tfigD}, and Fig.~\ref{tfigE}
  which parallel the analysis of
  Fig.~\ref{bfigB}, Fig.~\ref{bfigC}, Fig.~\ref{bfigD}, and Fig.~\ref{bfigE}.
  As in the b quark correction here also one finds that the SUSY QCD and 
  SUSY electroweak corrections to the top quark mass are strongly dependent 
  on the phases $\xi_3$, $\alpha_{A_0}$ and on the parameters 
  $m_0$, $m_{\frac{1}{2}}$ and $|A_0|$.  

\begin{figure}
\hspace*{-0.6in}
\centering
\includegraphics[width=12cm,height=16cm]{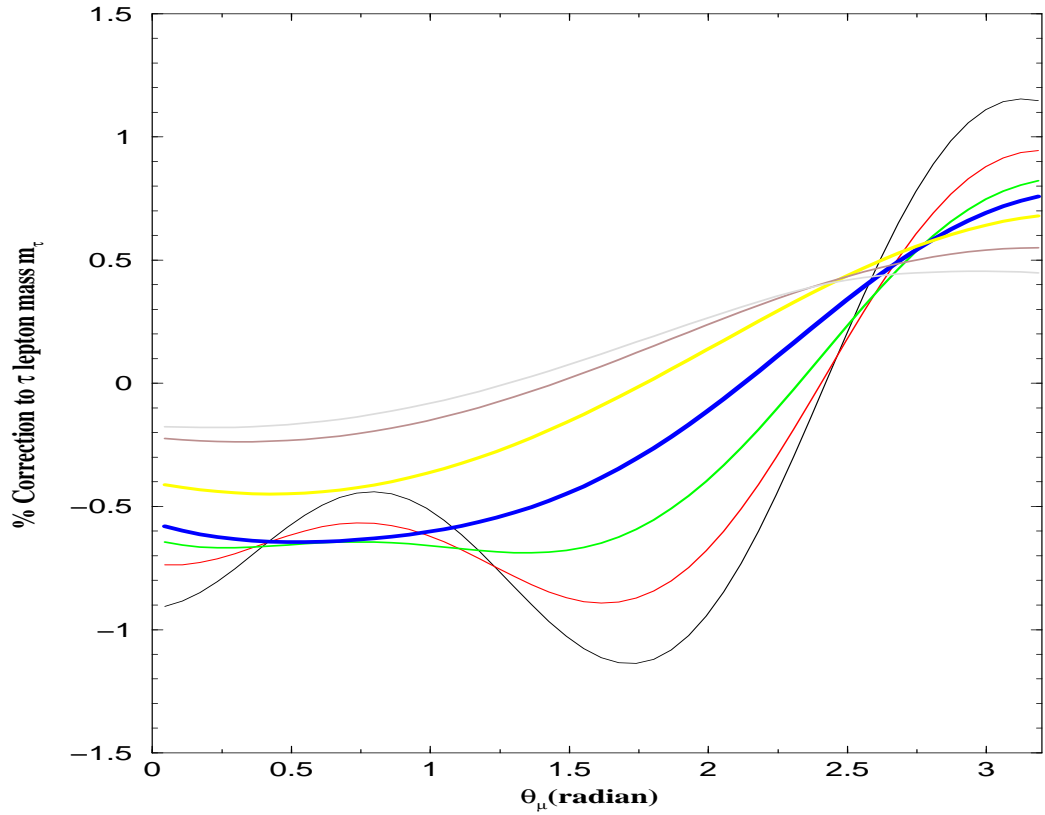}
\caption{Plot of the tau lepton mass correction $\Delta m_{\tau}/m_{\tau}$ 
as a function of $\theta_{\mu}$ for values of 
$\tan\beta$ ranging from 3 to 50. The other input parameters are:  
$m_0=m_{\frac{1}{2}}=200$ GeV,  $\xi_1=.5$, $\xi_2=.659$, 
$\xi_3=.633$, $\alpha_{A_0}=1.0$, and $|A_0|=4$.
The curves in ascending order at the point
$\theta_{\mu}= 0$ correspond to $\tan\beta =50,40,30,20,10,5,3$.
 }
\label{taufigA}
\end{figure}

\begin{figure}
\hspace*{-0.6in}
\centering
\includegraphics[width=12cm,height=16cm]{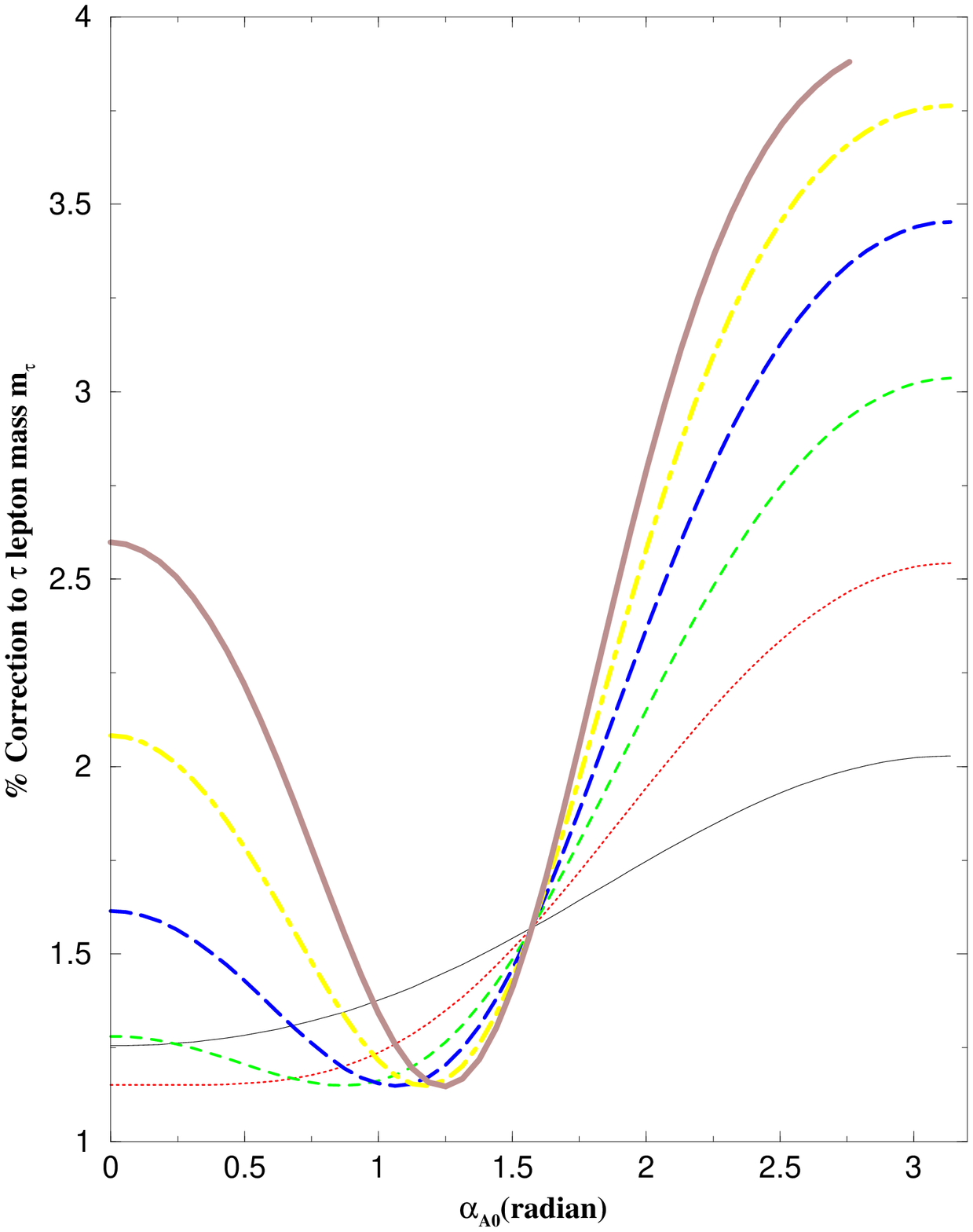}
\caption{Plot of the tau lepton mass correction $\Delta m_t/m_t$ 
in  percentage as a function of $\alpha_{A_0}$  
for various values of $|A_0|$. The other input parameters are:  
$m_0=m_{\frac{1}{2}}=200$ GeV,  $\xi_1=.5$, $\xi_2=.659$, 
$\xi_3=.633$, $\theta_{\mu}=\pi$ and $\tan\beta =50$.
The curves in ascending order at the point $ \alpha_{A_0}= 2.5$ 
correspond to  $|A_0|=1,2,3,4,5,6$}
\label{taufigD}
\end{figure}

\begin{figure}
\vspace{-2cm}
\hspace*{-0.6in}
\centering
\includegraphics[width=12cm,height=16cm]{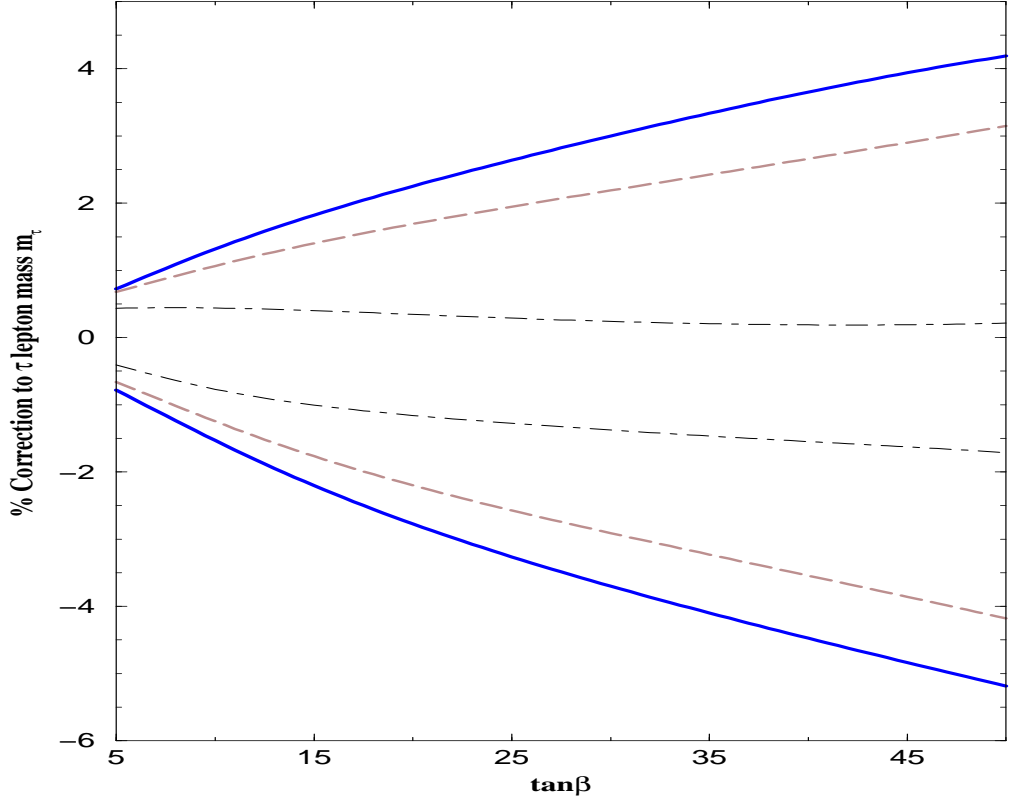}
\caption{Plot of the tau lepton mass correction $\Delta m_{\tau}/m_{\tau}$ 
in  percentage as a function of $\tan\beta$ for the three
cases (a), (b) and (c) of Table 1 of Ref.\cite{inhg199} where the EDM 
constraints for the electron, the neutron and $H_g^{199}$ atom are satisfied for 
the case $\tan\beta =50$. The upper dot-dashed curve is for the case 
(a) where $m_0=m_{\frac{1}{2}}=200$ GeV, $A_0=4$, $\alpha_{A_0}=1$,
  $\xi_1=.5$, $\xi_2=.659$, $\xi_3=.633$.
  The upper solid curve is for the case (b) where 
  $m_0=m_{\frac{1}{2}}=370$ GeV, $A_0=4$, $\alpha_{A_0}=2$,
  $\xi_1=.6$, $\xi_2=.653$, $\xi_3=.672$. 
  The upper long dashed curve is for the case (c) 
    where   $m_0=m_{\frac{1}{2}}=320$ GeV, $A_0=3$, $\alpha_{A_0}=.8$,
  $\xi_1=.4$, $\xi_2=.668$, $\xi_3=.6$. $\theta_{\mu}=2.5$ in all cases.
   The lower three similar
  curves have all the same parameters as cases (a), (b), and (c) above
  except that phases are set to zero.}
\label{taufigE}
\end{figure}

	Finally, we discuss the numerical size of the correction to the
 $\tau$ lepton mass.  In Fig.~\ref{taufigA} we give a plot of the  
  correction to the $\tau$ lepton mass  
  as a function of  $\theta_{\mu}$ for all the same parameters
  as in Fig.~\ref{bfigA} and in Fig.~\ref{tfigA}. 
  As in the case of the correction to the b quark mass and to the top 
  quark mass  
   here also one finds that the correction to the $\tau$ lepton 
  mass is very strongly dependent on  
  $\theta_{\mu}$. However, the numerical size of the correction is
  significantly smaller than the correction to the b quark mass 
  although generally bigger percentage wise than the percentage correction
  to the top quark mass. Again such effects  could be  relevant in
  precision physics. 
In Fig.~\ref{taufigD}, analogous to the case of Fig.~\ref{bfigD} for
the b quark and  Fig.~\ref{tfigD} for the case of the top quark, 
a plot of the $\tau$ lepton correction as a function of
$\alpha_{A_0}$ for various values of $|A_0|$ is given.
As in the case of  Fig.~\ref{bfigD} and Fig.~\ref{tfigD} one finds
a very strong dependence of the $\tau$ lepton mass correction 
on $\alpha_{A_0}$  and on $|A_0|$.  
In Fig.~\ref{taufigE}, as in the case of Fig.~\ref{bfigE} for the b quark
mass and in the case of Fig.~\ref{tfigE} for the top quark mass,
 we give a plot of the $\tau$ lepton correction as
a function of $\tan\beta$ for the three cases of Table 1 of
 Ref.\cite{inhg199} which satisfy the edm constraints for the 
 electron, the neutron and for the $H_g$ atom for value of 
 $\tan\beta =50$. The corresponding cases where the phases are all
 set to zero are also plotted. The plot shows that the effect of 
 phases consistent with the EDM constraints can produce large effects
 on the $\tau$ lepton  mass.  Fig.~\ref{taufigE} shows 
that the loop correction to the $\tau$ lepton mass can become
as large as 5\%. Thus overall we find that consistently in all cases
considered the supersymmetric QCD effects and the supersymmetric electroweak
effects are strongly dependent on CP phases. They can change both the
sign as well as the magnitude of the one loop correction. 
Such effects
will have important implications on low energy phenomena where 
supersymmetry QCD and supersymmetric electroweak effects enter.
An example of this is the analysis of branching ratio for the decay
$B^0_{s,d}\rightarrow l^+l^-$ decay which is strongly affected by
the SUSY QCD corrections\cite{inhg199,tata} and is also very sensitive 
to the CP phases.
As pointed out in Sec.1, Yukawa coupling unification is sensitive to 
SUSY QCD and SUSY electroweak loop correction and thus will also 
be sensitive to the CP phases.  

\section{Conclusion}
In this paper we have computed  the effects of CP  phases
on the $b$ and $t$  quark masses and the $\tau$ lepton mass. 
In Sec.2 we discussed the general technique for the computation of
QCD and electroweak loop corrections to the fermion masses and
specifically analysed the contributions to the b quark mass
arising from the gluino, chargino and neutralino exchange contributions.
We also checked that our results agree with previous analyses in
the limit when  phases are all set to zero and $\tan\beta$ is taken to be
large. In Sec.3 we extended the analysis to compute corrections to the
top quark mass arising from the gluino, chargino and neutralino exchange 
contributions. In Sec.4 we 
extended the analysis to compute the SUSY electroweak corrections to the 
$\tau$ lepton mass arising from chargino and neutralino exchange contributions.
These analyses were carried out allowing for the most general allowed
set of CP phases and without a constraint on $\tan\beta$. Thus the
analysis presented in this paper is valid not only for large $\tan\beta$ but
also for moderate and small values of $\tan\beta$. 
 The analysis presented in Secs.2, 3 and 4 is the first
complete one loop result of the corrections to the b and t  quark 
masses and to the $\tau$ lepton mass allowing for all allowed phases
and valid for all $\tan\beta$.  In Sec.5 we gave an exhaustive
numerical analysis of the size of the SUSY QCD and SUSY electroweak
corrections to the b and t  quark masses and to the $\tau$ lepton mass.
One finds that the numerical size of the corrections to the b quark
mass can be as large as 30\% while  the correction to the $\tau$ lepton
 mass as large as 5\%.  However, we find that the correction is typically less 
than a 1\% for the case of the top quark mass. Further, in all cases
analyzed one finds that the correction is sharply dependent on the
CP phases. Thus one finds that both the magnitude as well as the
sign may be affected by the presence of the CP phases. The loop
corrections are of great importance in unified models specifically
those involving Yukawa coupling unification, e.g., $b-\tau$ unification
for SU(5) and  $b-t-\tau$ unification for a class of SO(10) models.
More generally these loop corrections will be of import in the
study of quark-lepton textures in the presence of CP phases.
The SUSY QCD and SUSY electroweak corrections to the b quark mass is
also of great importance in the analysis of low energy phenomena
where the b quark mass enters prominently such as phenomena involving B mesons.
Thus, for example, the decay $B^0_{s,d}\rightarrow l^+l^-$ is strongly 
 affected by the SUSY QCD and SUSY electroweak 
 corrections\cite{inhg199,tata} .\\

\noindent
{\bf Acknowledgments}\\
\noindent
This research was supported in part by NSF grant PHY-0139967.\\

\newpage

\newpage

\newpage

\newpage

\newpage

\newpage

\newpage

\newpage

\newpage

\end{document}